# Multimillion Atom Simulations with NEMO 3-D

SHAIKH AHMED[2,1*], NEERAV KHARCHE[1*], RAJIB RAHMAN[1*], MUHAMMAD USMAN[1*], SUNHEE LEE[1*], HOON RYU[1], HANSANG BAE[1], STEVE CLARK[3], BENJAMIN HALEY[1], MAXIM NAUMOV[4], FAISAL SAIED[3], MAREK KORKUSINSKI[5], RICK KENNEL[3], MICHAEL MCLENNAN[3], TIMOTHY B. BOYKIN[6], AND GERHARD KLIMECK[1,7]

*Authors contributed equally

[1]School of Electrical and Computer Engineering and Network for Computational Nanotechnology, Purdue University, West Lafayette, IN 47907, USA.
[2]Electrical and Computer Engineering Department, Southern Illinois University, Carbondale, IL 62901, USA.
[3]Rosen Center for Advanced Computing, Purdue University, West Lafayette, IN 47907, USA.
[4]Department of Computer Science, Purdue University, West Lafayette, IN 47907, USA.
[5]Institute for Microstructural Sciences, National Research Council of Canada, 1200 Montreal Road, Ottawa, Ontario K1A 0R6.
[6]Electrical and Computer Engineering Dept., The University of Alabama in Huntsville, Huntsville, AL 35899.
[7]Jet Propulsion Laboratory, California Institute of Technology, Pasadena, CA 91109.

Tel: (618) 303 1439, Fax: (618) 453 7972, E-mail: ahmed@siu.edu
Tel: (765) 494 9212, Fax: (765) 494 6441, E-mail: gekco@purdue.edu

ARTICLE OUTLINE

## GLOSSARY AND NOTATION

*Nanostructures* — Nanostructures have at least two physical dimensions of size less than 100 nm. Their size lies between atomic/molecular and microscopic structures/particles. Realistically sized nanostructures are usually composed of millions of atoms. These devices demonstrate new capabilities and functionalities where the quantum nature of charge carriers plays an important role in determining the overall device properties and performance.

*Quantum dots* — Quantum dots (QDs) are solid-state nanostructures that provide confinement of charge carriers (electrons, holes, excitons) in all three spatial dimensions typically on the nanometer scale. This work focuses on semiconductor based quantum dots.

*Atomistic simulation* — For device sizes in the range of tens of nanometers, the atomistic granularity of constituent materials cannot be neglected. Effects of atomistic strain, surface roughness, unintentional doping, the underlying crystal symmetries, or distortions of the crystal lattice can have a dramatic impact on the device operation and performance. In an atomistic simulation, one takes into account both the atomistic/granular and quantum properties of the underlying nanostructure.

*Strain* — Strain is the deformation caused by the action of stress on a physical body. In nanoelectronic devices, strain typically originates from the assembly of lattice-mismatched semiconductors. Strain can be atomistically inhomogeneous and a small mechanical distortion of $2-5\%$ can strongly modify the energy spectrum, in particular the optical bandgap, of the system by $30-100\%$.

*Band structure* — Band structure of a solid originates from the wave nature of particles and depicts the allowed and forbidden energy states of electrons in the material. The knowledge of the band structure is the first and essential step towards the understanding of the device operation and reliable device design for semiconductor devices. Bandstructure is based on the assumption of an infinitely extended (bulk) material without spatial fluctuations (outside a simple repeated unit cell). For nanometer scale devices with spatial variations on the atomic scale the traditional concept of bandstructure is called into question.

*Piezoelectricity* — A variety of advanced materials of interest, such as GaAs, InAs, GaN, are piezoelectric. Piezoelectricity arises due to charge imbalances on the bonds between atoms. Modifications of the bond angles or distances result in alterations in charge imbalance. Any spatial non-symmetric distortion/strain in nanostructures made of these materials will create piezoelectric fields, which may significantly modify the electrostatic potential landscape.

*Tight binding* — Tight binding is an empirical model that enables calculation of single-particle energies and wave functions in a solid. The essential idea is the representation of the electronic states of the valence electrons with a local basis that contains the critical physical elements needed. The basis may contain orthogonal *s*, *p*, *d* orbitals on one atom that connect/talk to orbitals of a neighboring atom. The connection between atoms and the resulting overlapping wavefunctions form the bandstructure of a solid.

*NEMO 3-D* — NEMO 3-D stands for NanoElectronic MOdeling in three dimensions. This versatile, open source software package currently allows calculating single-particle electronic states and optical response of various semiconductor structures including bulk materials, quantum dots, impurities, quantum wires, quantum wells and nanocrystals.

*nanoHUB* — The nanoHUB is a rich, web-based resource for research, education and collaboration in nanotechnology (http://www.nanoHUB.org). It was created by the NSF-funded Network for Computational Nanotechnology (NCN) with a vision to pioneer the development of nanotechnology from science to manufacturing through innovative theory, exploratory simulation, and novel cyberinfrastructure. The nanoHUB offers online nanotechnology simulation tools which one can freely access from his/her web browser.

*Rappture* — Rappture (http://www.rappture.org) is a software toolkit that supports and enables the rapid development of graphical user interfaces (GUIs) for different applications. It is developed by Network for Computational Nanotechnology at Purdue University, West Lafayette.

I  DEFINITION OF THE SUBJECT AND ITS IMPORTANCE

The rapid progress in nanofabrication technologies has led to the emergence of new classes of nanodevices and structures which are expected to bring about fundamental and revolutionary changes in electronic, photonic, computation, information processing, biotechnology, and medical industries. At the atomic scale of novel nanostructured semiconductors the distinction between new device and new material is blurred and device physics and material science meet. The *quantum mechanical effects* in the electronic states of the device and the *granular, atomistic* representation of the underlying material become important. Modeling and simulation approaches based on a *continuum* representation of the underlying material typically used by device engineers and physicists become invalid. Typical *ab initio* methods used by material scientists do not represent the bandgaps and masses precisely enough for device design or they do not scale to *realistically sized devices which may contain millions of atoms*. The variety of geometries, materials, and doping configurations in semiconductor devices at the nanoscale suggests that a *general* nanoelectronic modeling tool is needed. The Nanoelectronic Modeling tool (NEMO 3-D) has been developed to address these needs. Based on the atomistic valence-force field (VFF) method and a variety of nearest-neighbor tight-binding models ($s$, $sp^3s^*$, $sp^3d^5s^*$), NEMO 3-D enables the computation of strain for over 64 million atoms and of electronic structure for over 52 million atoms, corresponding to volumes of $(110nm)^3$ and $(101nm)^3$, respectively. Such extreme problem sizes involve very large-scale computations, and NEMO 3-D has been designed and optimized to be scalable from single CPUs to large numbers of processors on commodity clusters and the most advanced supercomputers. Excellent scaling to 8192 cores/CPUs has been demonstrated. NEMO 3-D is continually developed by the Network for Computational Nanotechnology (NCN) under an open source license. A web-based online interactive version for educational purposes is freely available on the NCN portal http://www.nanoHUB.org. This article discusses the theoretical models, essential algorithmic and computational components, and optimization methods that have been used in the development and the deployment of NEMO 3-D. Also, successful applications of NEMO 3-D are demonstrated in the atomistic calculation of single-particle electronic states of the following realistically-sized nanostructures each consisting of multimillion atoms: (1) self-assembled quantum dots including long-range strain and piezoelectricity; (2) stacked quantum dots as used in quantum cascade lasers; (3) Phosphorus (P) impurities in Silicon used in quantum computation; (4) Si on SiGe quantum wells (QWs); and (5) SiGe nanowires. These examples demonstrate the broad NEMO 3-D capabilities and indicate the necessity of multimillion atomistic electronic structure modeling.

II  INTRODUCTION

*(A) Emergence of Novel Nanoscale Semiconductor Devices*

The new industrial age and the new economy are driven in large measure by unprecedented advances in information technology. The electronics industry is the largest industry in the world with global sales of over one trillion dollars since 1998. If current trends continue, the sales volume of the electronics industry is predicted to reach three trillion dollars and account for about 10% of gross world product (GWP) by 2010 [93]. Basic to the electronic industry and the new information age are the *semiconductor devices* that implement all needed information processing operations. The revolution in the semiconductor industry was initiated in

1947 with the invention and fabrication of point-contact bipolar devices on *slabs* of polycrystalline germanium (Ge) used as the underlying semiconductor element [1]. Later the development of the planar process and the reliable and high-quality silicon dioxide ($SiO_2$) growth on silicon wafers, acting as an excellent barrier for the selective diffusion steps, led to the invention of the silicon-based bipolar integrated circuits in 1959. A metal–oxide–semiconductor field-effect transistor (MOSFET), the most critical device for today's advanced integrated circuits, was reported by Kahng and Atalla in 1960 [93]. By 1968, both complementary metal–oxide–semiconductor devices (CMOS) and polysilicon gate technology allowing self-alignment of the gate to the source/drain of the device had been developed. The industry's transition from bipolar to CMOS technology in the 1980s was mainly driven by the increased power demand for high-performance integrated circuits.

The most important factor driving the continuous device improvement has been the semiconductor industry's relentless effort to reduce the cost *per function* on a chip [96]. This is done by putting more devices on an integrated circuit chip while either reducing manufacturing costs or holding them constant. *Device scaling*, which involves reducing the transistor size while keeping the electric field constant from one generation to the next, has paved the way for a continuous and systematic increase in transistor density and improvements in system performance (described by Moore's Law [69]) for the past forty years. For example, regarding conventional/classical silicon MOSFETs, the device size is scaled in all dimensions, resulting in smaller oxide thickness, junction depth, channel length, channel width, and isolation spacing. Currently, 65 nm (with a physical gate length of 35 nm) is the state-of-the-art process technology, but even smaller dimensions are expected in the very near future.

However, recent studies by many researchers around the globe reveal the fact that the exponential growth in integrated circuit complexity as achieved through conventional scaling is finally facing its limits and will slow down in very near future. Critical dimensions, such as transistor gate length and oxide thickness, are reaching physical limitations [96]. Maintaining dimensional integrity at the limits of scaling is a challenge. Considering the manufacturing issues, photolithography becomes difficult as the feature sizes approach the wavelength of ultraviolet light. In addition, it is difficult to control the oxide thickness when the oxide is made up of just a few monolayers. Processes will be required approaching atomic-layer precision. In addition to the processing issues there are also some fundamental device issues [103]. As the silicon industry moves into the 45 nm node regime and beyond, two of the most important challenges facing us are the growing dissipation of *standby power* and the increasing variability and mismatch in device characteristics.

The Semiconductor Industry Association (SIA) forecasts [88] that the current rate of transistor performance improvement can be sustained for another 10 to 15 years, but only through the development and introduction of *new materials and transistor structures*. In addition, a major improvement in lithography will be required to continue size reduction. It is expected that these new technologies may extend MOSFETs to the 22 nm node (9-nm physical gate length) by 2016. Intrinsic device speed may exceed 1 THz and integration densities will be more than 1 billion transistors/$cm^2$. In many cases, the introduction of a new material requires the use of a new device structure, or vice versa. To fabricate devices beyond current scaling limits, IC companies are simultaneously pushing the planar, bulk silicon CMOS design while exploring alternative gate stack materials (high-*k* dielectric [108] and metal gates), band engineering methods (using strained Si [102] or SiGe [72]), and alternative transistor structures. The concept of a band-engineered transistor is to enhance the mobility of electrons and/or holes in the channel by modifying the band structure of silicon in the channel in a way such that the physical structure of the transistor remains substantially unchanged. This enhanced mobility increases the transistor transconductance ($g_m$) and on-drive current ($I_{on}$). A SiGe layer or a strained-silicon on relaxed SiGe layer is used as the enhanced-mobility channel layer. Today there is also an extensive research in double-gate (DG) structures, and FinFET transistors [23], which have better electrostatic integrity and theoretically have better transport properties than single-gated FETs. Some novel and revolutionary technology such as carbon nanotubes, silicon nanowires, or molecular transistors might be seen on the horizon, but it is not obvious, in view of the predicted future capabilities of CMOS, how competitive they will be.

A recent analysis based on fundamental quantum mechanical principles, restated by George Bourianoff of the Intel Corporation, reveals that heat/power dissipation will ultimately limit any logic device using an electronic charge [107] and operating at room temperature. This limit is about 100 watts per square centimeter for passive cooling techniques with no active or electrothermal elements. These fundamental limits have led to pessimistic predictions of the imminent end of technological progress for the semiconductor industry and simultaneously have increased interest in advanced alternative technologies that rely on something other than electronic charge—such as spin or photon fields—to store computational state. Many advocate a focus on quantum computers that make use of distinctively quantum mechanical phenomena, such as entanglement and superposition, to perform operations on data. Among a number of quantum computing proposals, the Kane scalable quantum computer is based on an array of individual phosphorus (P) donor atoms embedded in a pure silicon lattice [41]. Both the nuclear spins of the donors and the spins of the donor electrons participate in the quantum computation. The Loss-DiVincenzo quantum computer [63], also a scalable semiconductor-based quantum computer, makes use of the intrinsic spin degree of freedom of individual electrons confined to quantum dots as qubits.

TABLE I

MAJOR SEMICONDUCTOR DEVICES WITH THE APPROXIMATE DATE OF THEIR INTRODUCTION

- 1874: Metal-semiconductor contact
- 1947: Bipolar junction transistors (BJT)
- 1954: Solar cell
- 1957: Heterojunction bipolar transistor (HBT)
- 1958: Tunnel diode
- 1959: Integrated circuits
- 1960: Field-effect transistors (FETs)
- 1962: Semiconductor lasers.
- 1966: Metal-semiconductor FET
- 1967: Nonvolatile semiconductor memory
- 1974: Resonant tunneling diode (RTD)
- 1990: Magnetoresistive Random Access Memory (MRAM)
- 1991: Carbon nanotubes
- 1994: Room-temperature single-electron memory cell (SEMC)
- 1994: Quantum Cascade Laser
- 1998: Carbon nanotube FET
- 1998: Proposal for Kane quantum computer
- 2001: 15 nm MOSFET
- 2003: High performance Silicon nanowire FET

Since the invention of the point-contact bipolar transistor in 1947, advanced fabrication technologies, introduction of new materials with unique properties, and broadened understanding of the underlying physical processes have resulted in tremendous growth in the number and variety of semiconductor devices and literally changed the world. To date, there are about 60 major devices, with over 100 device variations related to them. A list of most of the basic semiconductor devices (mainly based on Ref. [93]) discovered and used over the past century with the date of their introduction is shown in Table 1.

### (B) Need for Simulations

Simulation is playing key role in device development today. Two issues make simulation important [96]. Product cycles are getting shorter with each generation, and the demand for production wafers shadows development efforts in the factory. Consider the product cycle issue first. In order for companies to maintain their competitive edge, products have to be taken from design to production in less than 18 months. As a result, the development phase of the cycle is getting shorter. Contrast this requirement with the fact that it takes 2–3 months to run a wafer lot through a factory, depending on its complexity. The specifications for experiments run through the factory must be near the final solution. While simulations may not be completely predictive, they provide a good initial guess. This can ultimately reduce the number of iterations during the device development phase.

The second issue that reinforces the need for simulation is the production pressures that factories face. In order to meet customer demand, development factories are making way for production space. It is also expensive to run experiments through a production facility. The displaced resources could have otherwise been used to produce sellable product. Again, device simulation can be used to decrease the number of experiments run through a factory. Device simulation can be used as a tool to guide manufacturing down a more efficient path, thereby decreasing the development time and costs.

Besides offering the possibility to test hypothetical devices which have not (or could not have) yet been manufactured, device simulation offers unique insight into device behavior by allowing the observation of internal phenomena that can not be measured. Thus, a critical facet of the nanodevices development is the creation of simulation tools that can quantitatively explain or even predict experiments. In particular it would be very desirable to explore the design space before, or in conjunction with, the (typically time consuming and expensive) experiments. A general tool that is applicable over a large set of materials and geometries is highly desirable. But the tool development itself is not enough. The tool needs to be deployed to the user community so it can be made more reliable, flexible, and accurate.

### (C) Goal of This Article

The rapid progress in nanofabrication technologies has led to the development of novel devices and structures which could revolutionize many high technology industries. These devices demonstrate new capabilities and functionalities where the *quantum nature* of charge carriers plays an important role in determining the overall device properties and performance. For device sizes in the range of tens of nanometers, the *atomistic granularity* of constituent materials cannot be neglected: effects of atomistic strain, surface roughness, unintentional doping, the underlying crystal symmetries, or distortions of the crystal lattice can have a dramatic impact on the device operation and performance.

The goal of this paper is to describe the theoretical models and the essential algorithmic and computational components that have been used in the development and deployment of the Nanoelectronic Modeling tool NEMO 3-D on http://www.nanoHUB.org and to demonstrate successful applications of NEMO 3-D in the atomistic calculation of single-particle electronic states of different, realistically sized nanostructures, each consisting of multi-million atoms. We present some of the new capabilities that have been recently added to NEMO 3-D to make it one of the premier simulation tools for design and analysis of realistic nanoelectronic devices, and thus a valid tool for the computational nanotechnology community. These recent advances include algorithmic refinements, performance analysis to identify the best computational strategies, and memory saving measures. The effective scalability of NEMO 3-D code is demonstrated on the IBM BlueGene, the Cray XT3, an Intel Woodcrest cluster, and other Linux clusters. The largest electronic structure calculation, with *52 million atoms,* involved a Hamiltonian matrix with over one *billion* complex degrees of freedom. The performance impact of storing the Hamiltonain versus recomputing the matrix, when needed, is explored. We describe the state-of-the-art algorithms that have been incorporated in the code, including very effective Lanczos, block Lanczos and Tracemin eigenvalue solvers, and present a comparison of the different solvers. While system sizes of tens of millions of atoms appear at first sight huge and wasteful, we demonstrate that some physical problems require such large scale analysis. We recently showed [44] that the analysis of valley splitting in strained Si quantum wells grown on strained SiGe required atomistic analysis of 10 million atoms to match experimental data. The insight that disorder in the SiGe buffer increases valley splitting in the Si quantum well would probably not be predictable in a continuum effective mass model. Similarly, the simulations of P impurities in silicon required multi-million atom simulations [82]. In the following, we describe NEMO 3-D capabilities in the simulation of different classes of nanodevices having carrier confinement in 3, 2, and 1 dimensions in the GaAs/InAs and SiGe materials systems.

*Single and Stacked Quantum Dots* (*confinement in 3 dimensions*). Quantum dots (QDs) are solid-state semiconducting nanostructures that provide confinement of charge carriers (electrons, holes, excitons) in all three spatial dimensions resulting in strongly localized wave functions, discrete energy eigenvalues and interesting physical and novel device properties [68][85][84][6][77][70]. Existing nanofabrication techniques tailor QDs in a variety of types, shapes and sizes. Within bottom-up approaches, QDs can be realized by colloidal synthesis at benchtop conditions. Quantum dots thus created have dimensions ranging from 2–10 nanometers, corresponding to 100–100,000 atoms.

Self-assembled quantum dots (SAQDs) grown in the coherent Stranski-Krastanov heteroepitaxial growth mode nucleate spontaneously within a lattice mismatched material system (for example, InAs grown on GaAs substrate) under the influence of strain in certain physical conditions during molecular beam epitaxy (MBE) and metalorganic vapor phase epitaxy (MOVPE) [3]. The strain produces coherently strained quantum-sized islands on top of a two-dimensional wetting-layer. The islands can be subsequently buried. Semiconducting QDs grown by self-assembly are of particular importance in quantum optics [28][67], since they can be used as detectors of infrared radiation, optical memories, and in laser applications.

The strongly peaked energy dependence of density of states and the strong overlap of spatially confined electron and hole wavefunctions provide ultra-low laser threshold current densities, high temperature stability of the threshold current, and high material and differential quantum gain/yield. Strong oscillator strength and non-linearity in the optical properties have also been observed [67]. Self-assembled quantum dots also have potential for applications in quantum cryptography as single photon sources and quantum computation [41][22]**.** In electronic applications QDs have been used to operate like a single-electron transistor and demonstrate a pronounced Coulomb blockade effect. Self-assembled QDs, with an average height of 1–5 nm, are typically of size (base length/diameter) 5–50 nm and consist of 5,000–2,000,000 atoms. Arrays of quantum-mechanically coupled (stacked) self-assembled quantum dots can be used as optically active regions in high-efficiency, room-temperature lasers. Typical QD stacks consist of 3–7 QDs with typical lateral extension of 10–50 nm and dot height of 1–3 nm. Such dots contain 5–50 million atoms in total, where atomistic details of interfaces are extremely important [95].

*Impurities* (*confinement in 3 dimensions*). Impurities have always played a vital role in semiconductors since the inception of the transistor. Till the end of last century, scientists and engineers had been interested in the macroscopic properties of an ensemble of dopants in a semiconductor. As technology enters the era of nanoscale electronics, devices which contain a few discrete dopants are becoming increasingly common. In recent years, there have been proposals of novel devices that operate on purely quantum mechanical principles using the quantum states of isolated or coupled donors/impurities [41][97][36]. The on-going extensive research effort on the Phosphorus (P) donor based quantum computer architecture of Kane [41] exemplifies an effort to harness the quantum nature of materials for the development of next generation electronics. As researchers strive to establish atomic scale quantum control over single impurities [87][19][91], precision modeling techniques are required to explore this new regime of device operations [25][65][82][29].

Although effective mass based approaches have been predominantly used in literature to study the physics of impurities, realistic device modeling using this technique have proved difficult in practice. Tight-binding methods [89] consider a more extensive Bloch structure for the host material, and can treat interfaces, external gates, strain, magnetic fields, and alloy disorder within a single framework. When applied to realistic nanodevices of several million atoms, this technique can prove very effective for device modeling [50]. We present a semi-empirical method for modeling impurities in Si that can be used for a variety of applications such as *quantum computer* architecture, discretely doped FinFETs, and impurity scattering problems. Although we focus on P impurities in Si here, the method is sufficiently general to be used on other impurities and hosts.

*Quantum Wires* (*confinement in 2 dimensions*). For quite some time, nanowires have been considered a promising candidate for future building block in computers and information processing machines [49][98][106][64][8]. Nanowires are fabricated from different materials (metal, semiconductor, insulator and molecular) and assume different cross-sectional shapes, dimensions and diameters. Electrical conductivity of nanowires is greatly influenced by edge effects on the surface of the nanowire and is determined by quantum mechanical conductance. In the nanometer regime, the impact of surface roughness or alloy disorder on electronic bandstructure must be atomistically studied to further gauge the transport properties of nanowires.

*Quantum Wells* (*confinement in 1 dimension*). QW devices are already a de-facto standard technology in MOS devices and QW lasers. They continue to be examined carefully for ultra-scaled devices where interfacial details turn out to be critical. Composite channel materials with GaAs, InAs, InSb, GaSb, and Si are being considered [81][78], which effectively constitute QWs. Si QWs buffered/strained by SiGe are considered for Quantum Computing (QC) devices where valley-splitting (VS) is an important issue [27]. Si is desirable for QC due to its long spin-decoherence times, scaling potential and integrability within the present microelectronic manufacturing infrastructure. In strained Si, the 6-fold valley-degeneracy of Si is broken into lower 2-fold and raised 4-fold valley-degeneracies. The presence of 2-fold valley-degeneracy is a potential source of decoherence which leads to leakage of quantum information outside qubit Hilbert space. Therefore, it is of great interest to study the lifting of the remaining 2-fold valley degeneracy in strained Si due to sharp confinement potentials in recently proposed [27] SiGe/Si/SiGe quantum well (QW) heterostructures based quantum computing architectures.

## III NANOSCALE DEVICE MODELING AND SIMULATION CHALLENGES

The theoretical knowledge of the electronic structure of nanoscale semiconductor devices is the first and most essential step towards the interpretation and the understanding of the experimental data and reliable device design at the nanometer scale. The following is a list of the modeling and simulation challenges in the design and analysis of realistically sized engineered nanodevices.

(1) *Full Three-Dimensional Atomistic Representation*: The lack of *spatial symmetry* in the overall geometry of the nanodevices usually requires explicit three-dimensional representation. For example, Stranski-Krastanov growth techniques tend to produce self-assembled InGaAs/GaAs quantum dots [68][85][84] with some rotational symmetry, e.g. disks, truncated cones, domes, or pyramids [6]. These structures are generally not perfect geometric objects, since they are subject to interface interdiffusion, and discretization on an atomic lattice. There is no such thing as a round disk on a crystal lattice! The underlying crystal symmetry imposes immediate restrictions on the realistic geometry and influences the quantum mechanics. Continuum methods such as effective mass [78] and $k \bullet p$ [35][92] typically ignore such crystal symmetry and atomistic resolution.

The required simulation domain sizes of ~1M atoms prevent the usage of *ab initio* methods. Empirical methods which eliminate enough unnecessary details of core electrons, but are finely tuned to describe the atomistically dependent behavior of valence and conduction electrons, are needed. The current state-of-the-art leaves 2 choices: 1) pseudopotentials [20] and 2) Tight Binding [50]. Both methods have their advantages and disadvantages. Pseudopotentials use plane waves as a fundamental basis choice. Realistic nanostructures contain high frequency features such as alloy-disorder or hetero-interfaces. This means that the basis needs to be adjusted (by an expert) for every different device, which limit the potential impact for non-expert users. Numerical implementations of pseudopotential calculations typically require a Fourier transform between real and momentum space which demand full matrix manipulations and full transposes. This typically requires high bandwidth communication capability (i.e. extremely expensive) parallel machines, which limit the practical dissemination of the software to end users with limited compute resources. Tight-binding is a local basis representation, which naturally deals with finite device sizes, alloy-disorder and hetero-interfaces and it results in very sparse matrices. The requirements of storage and processor communication are therefore minimal compared to pseudopotentials and actual implementations perform extremely well on inexpensive clusters [50].

Tight-binding has the disadvantage that it is based on empirical fitting and some in the community continue to question the fundamental applicability of tight-binding. The NEMO team has spent a significant effort to expand and document the tight-binding capabilities with respect to handling of strain [14], electromagnetic fields [10], and Coulomb matrix elements [59] and fit them to well known and accepted bulk parameters [50][47][46]. With tight-binding the NEMO team was able early on to match experimentally verified, high-bias current-voltage curves of resonant tunneling [7][48] that could not get modeled by ether effective mass (due to the lack of physics) or pseudopotential methods (due to the lack of open boundary conditions). We continue to learn about the tight-binding method capabilities, and we are in the process of benchmarking it against more fundamental *ab initio* approaches and pseudopotential approaches. Our current Si/Ge parameterization is described in references [15][11]. Figure 1 depicts a range of phenomena that represent new challenges presented by new trends in nanoelectronics and lays out the NEMO 3-D modeling agenda.

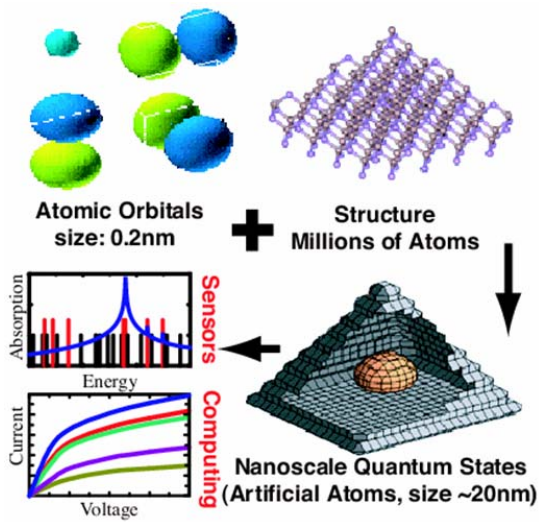

**Figure 1**. NEMO 3-D modeling agenda: map electronic properties of individual atoms into realistic structures containing millions of atoms, computation of nanoscale quantum dots that maps into real applications.

(2) *Atomistic Strain*: Strain that originates from the assembly of lattice-mismatched semiconductors strongly modifies the energy spectrum of the system. In the case of the InAs/GaAs quantum dots, this mismatch is around 7% and leads to a strong *long-range* strain field within the extended neighborhood (typically ~ 25 nm) of each quantum dot [2]. Si/Ge core/shell structured nanowires are another example of strain dominated atom arrangements [62]. Si quantum wells and SiGe quantum computing architectures rely on strain for state separation [27]. The strain can be atomistically inhomogeneous, involving not only biaxial components but also non-negligible shear components. Strain strongly influences the core and barrier material band structures, modifies the energy bandgaps, and lifts the heavy hole-light hole degeneracy at the zone center. In the nanoscale regime, the classical harmonic linear/continuum elasticity model for strain is inadequate, and device simulations must include the fundamental quantum character of charge carriers and the long-distance atomistic strain effects with proper boundary conditions on equal footing [101][58].

(3) *Piezoelectric Field*: A variety of III-IV materials such as GaAs, InAs, GaN, are piezoelectric. Any spatial non-symmetric distortion in nanostructures made of these materials will create piezoelectric fields, which will modify the electrostatic potential landscape. Recent spectroscopic analyses of self-assembled QDs demonstrate polarized transitions between confined hole and electron levels [6]. While the continuum models (effective mass or $k \bullet p$) can reliably predict aspects of the single-particle energy states, they fail to capture the observed non-degeneracy and optical polarization anisotropy of the excited energy states in the (001) plane. These methods fail because they use a confinement potential which is assumed to have only the *shape symmetry* of the nanostructure, and they ignore the underlying crystal symmetry. The experimentally measured symmetry is significantly lower than the assumed continuum symmetry because of (a) underlying crystalline symmetry, (b) atomistic strain relaxation and (c) piezoelectric field. For example, in the case of pyramid shaped quantum dots with square bases, continuum models treat the underlying material in $C_{4v}$ symmetry while the atomistic representation lowers the crystal symmetry to $C_{2v}$. The piezoelectric potential originating from the non-zero shear component of the strain field must be taken into account to properly model the associated symmetry breaking and the introduction of a global shift in the energy spectra of the system.

## IV    NEMO 3-D SIMULATION PACKAGE

### *(A) Basic Features — Simulation Domains*

NEMO 3-D [50][53][74][75][55] bridges the gap between the large size, classical semiconductor device models and the molecular level modeling. This package currently allows calculating single-particle electronic states and optical response of various semiconductor structures including bulk materials, quantum dots, quantum wires, quantum wells and nanocrystals. NEMO 3-D includes spin in its fundamental atomistic tight binding representation. Spin is therefore not added in as an afterthought into the theory, but spin-spin interactions are naturally included in the Hamiltonian. Effects of interaction with external electromagnetic fields are also included [50][31][10]. A schematic view of InAs quantum dot embedded in a GaAs barrier material the sample is presented in Figure 2. The quantum dot is positioned on a 0.6 nm thick wetting layer (dark region). The simulation of strain is carried out in the large computational box $D_{strain}$, while the electronic structure computation is restricted to the smaller domain $D_{elec}$. Strain is long-ranged and penetrates around 25 nm into the dot substrate thus stressing the need for using large substrate thickness in the simulations. NEMO 3-D enables the computation of strain and electronic structure in an atomistic basis for over 64 and 52 million atoms, corresponding to volumes of $(110nm)^3$ and $(101nm)^3$, respectively. These volumes can be spread out arbitrarily over any closed geometry. For example, if a thin layer of 15 nm height is considered, the corresponding widths in the x-y plane correspond to 298 nm for strain calculations and 262 nm for electronic structure calculations. No other atomistic tool can currently handle such volumes needed for realistic device simulations. NEMO 3-D runs on serial and parallel platforms, local cluster computers as well as the NSF Teragrid.

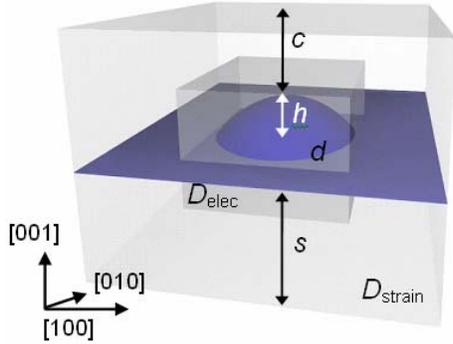

**Figure 2**. Simulated dome shaped InAs quantum dot buried in GaAs. Two simulation domains are shown, $D_{elec}$: central smaller domain for electronic structure calculation, and $D_{strain}$: outer larger domain for strain calculation. In the figure: s is the substrate height, c is the cap layer thickness, h is the dot height, d is the dot diameter.

*(B) Components and Models*

The NEMO 3-D program flow consists of four main components.

(1) *Geometry Construction*. The first part is the geometry constructor, whose purpose is to represent the treated nanostructure in atomistic detail in the memory of the computer. Each atom is assigned three single-precision numbers representing its coordinates, stored is also its type (atomic number in short integer), information whether the atom is on the surface or in the interior of the sample (important later on in electronic calculations), what kind of computation it will take part of (strain only or strain and electronic), and what its nearest neighbor relation in a unit cell is. The arrays holding this structural information are initialized for all atoms on all CPUs, i.e., the complete information on the structure is available on each CPU. By default most of this information can be stored in short integer arrays or as single bit arrays, which does not require significant memory. This serial memory allocation of the atom positions, however, becomes significant for very large systems which must be treated in parallel.

(2) *Strain*. The materials making up the QD nanostructure may differ in their lattice constants; for the InAs/GaAs system this difference is of the order of 7%. This lattice mismatch leads to the appearance of strain: atoms throughout the sample are displaced from their bulk positions. Knowledge of equilibrium atomic positions is crucial for the subsequent calculation of QD's electronic properties, which makes the computation of strain a necessary step in realistic simulations of these nanostructures.

NEMO 3-D computes strain field using an atomistic valence force field (VFF) method [42] with the Keating Potential. In this approach, the total elastic energy of the sample is computed as a sum of bond-stretching and bond-bending contributions from each atom. The local strain energy at atom $i$ is given by a phenomenological formula

$$E_i = \frac{3}{8} \sum_j \left[ \frac{\alpha_{ij}}{2d_{ij}^2} \left(R_{ij}^2 - d_{ij}^2\right)^2 + \sum_{k>j}^n \frac{\sqrt{\beta_{ij}\beta_{ik}}}{d_{ij}d_{ik}} \left(\vec{R}_{ij} \cdot \vec{R}_{ik} - \vec{d}_{ij} \cdot \vec{d}_{ik}\right)^2 \right], \quad (1)$$

where the sum is carried out over the $n$ nearest neighbors $j$ of atom $i$, $\vec{d}_{ij}$ and $\vec{R}_{ij}$ are the bulk and actual (distorted) distances between neighbor atoms, respectively, and $\alpha_{ij}$ and $\beta_{ij}$ are empirical material-dependent elastic parameters. The equilibrium atomic positions are found by minimizing the total elastic energy of the system. Several other strain potentials [101] [58] are also implemented in NEMO 3-D. While they modify some of the strain details they roughly have the same computational efficiency.

(3) *Electronic Structure*. The single-particle energies and wave functions are calculated using an empirical nearest-neighbor tight-binding model. The underlying idea of this approach is the selection of a basis consisting of atomic orbitals (such as $s$, $p$, $d$, and $s^*$) centered on each atom. These orbitals are further treated as a basis set for the Hamiltonian, which assumes the following form:

$$\hat{H} = \sum_i \varepsilon_i^{(\nu)} c_{i,\nu}^+ c_{i,\nu} + \sum_{i,\nu,\mu} t_i^{(\nu\mu)} c_{i,\nu}^+ c_{i,\mu} + \sum_{i,j,\nu,\mu} t_{ij}^{(\nu\mu)} c_{i,\nu}^+ c_{j,\mu}, \quad (2)$$

where $c_{i,\nu}^+$ ($c_{i,\nu}$) is the creation (annihilation) operator of an electron on the orbital $\nu$ localized on atom $i$. In the above equation, the first term describes the onsite orbital terms, found on the diagonal of the Hamiltonian matrix. The second term describes coupling between different orbitals localized on the same atom (only the spin-orbit coupling between $p$-orbitals), and the third term describes coupling between different orbitals on different atoms. The restriction in the summation of the last term is that the atoms $i$ and $j$ be nearest neighbors.

The characteristic parameters $\varepsilon$ and $t$ are treated as empirical fitting parameters for each constituent material and bond type. They are usually expressed in terms of energy constants of $\sigma$ and $\pi$ bonds between the atomic orbitals. For example, for a simple cubic lattice, the interaction between the $s$ orbital localized on the atom $i$ at origin and the orbital $p_x$ localized on the atom $j$ with coordinate $\vec{d}_{ij} = a\hat{x}$ with respect to the atom $i$ would simply be expressed as $t_{ij}^{(s,p_x)} = V_{sp\sigma}$. Most of the systems under consideration, however, crystallize in the zinc-blende lattice, which means that the distance between the nearest neighbors is described by a 3-D vector $\vec{d}_{ij} = l\hat{x} + m\hat{y} + n\hat{z}$, with $l$, $m$, $n$ being the directional cosines. These cosines rescale the interaction constants, so that the element describing the interaction of the orbitals $s$ and $p_x$ is $t_{ij}^{(s,p_x)} = lV_{sp\sigma}$. The parameterization of all bonds using analytical forms of directional cosines for various tight-binding models is given in Ref. [90]. NEMO 3-D provides the user with choices of the $sp^3d^5s^*$, $sp^3s^*$, and single $s$-orbital models with and without spin, in zincblende, wurzite, and simple cubic lattices.

Additional complications arise in strained structures, where the atomic positions deviate from the ideal (bulk) crystal lattice [40]. The presence of strain leads to distortions not only of bond directions, but also bond lengths. In this case, the discussed interaction constant $t_{ij}^{(s,p_x)} = l'V_{sp\sigma}\left(\dfrac{d}{d_0}\right)^{\eta(sp\sigma)}$, where the new directional cosine $l'$ can be obtained analytically from the relaxed atom positions, but the bond-stretch exponent $\eta(sp\sigma)$ needs to be fitted to available data. The energy constants parameterizing the on-site interaction change as well due to bond renormalization [50][14].

The 20-band nearest-neighbor tight-binding model is thus parameterized by 34 energy constants and 33 strain parameters, which need to be established by fitting the computed electronic properties of materials to those measured experimentally. This is done by considering bulk semiconductor crystals (such as GaAs or InAs) under strain. The summation in the Hamiltonian for these systems is done over the primitive crystallographic unit cell only. The model makes it possible to compute the band structure of the semiconductor throughout the entire Brillouin zone. For the purpose of the fitting procedure, however, only the band energies and effective masses at high symmetry points and along the $\Delta$ line from $\Gamma$ to X are targeted, and the tight-binding parameters are adjusted until a set of values closely reproducing these target values is found. Search for optimal parameterization is done using a genetic algorithm, described in detail in Refs. [50][31]. Once it is known for each material constituting the QD, a full atomistic calculation of the single-particle energy spectrum is carried out on samples composed of millions of atoms. No further material properties are adjusted for the nanostructure, once they are defined as basic bulk material properties.

(4) *Post Processing of Eigenstates.* From the single-particle eigenstates various physical properties can be calculated in NEMO 3-D such as optical matrix elements [9], Coulomb and exchange matrix elements [59], approximate single cell bandstructures from supercell bandstructure [13][12][8].

*(C) Algorithmic and Numerical Aspects*

(1) *Parallel Implementation.* The complexity and generality of physical models in NEMO 3-D can place high demands on computational resources. For example, in the 20-band electronic calculation the discrete Hamiltonian matrix is of order 20 times the number of atoms. Thus, in a computation with 20 million atoms, the matrix is of order 400 million. Computations of that size can be handled because of the parallelized design of the package. NEMO 3-D is implemented in ANSI C, C++ with MPI used for message-passing, which ensures its portability to all major high-performance computing platforms, and allows for an efficient use of distributed memory and parallel execution mechanisms.

Although the strain and electronic parts of the computation are algorithmically different, the key element in both is the sparse matrix-vector multiplication. This allows the use of the same memory distribution model in both phases. The computational domain is divided into slabs along one dimension. All atoms from the same slab are assigned to a single CPU, so if all nearest neighbors of an atom belong to its slab, no inter-CPU communication is necessary. The interatomic couplings are then fully contained in one of the diagonal blocks of the matrix. On the other hand, if an atom is positioned on the interface between slabs, it will couple to atoms belonging both to its own and the neighboring slab. This coupling is described by the off-diagonal blocks of the matrix. Its proper handling requires inter-CPU communication. However, due to the first-nearest-neighbor character of the strain and electronic models, the messages need to be passed only between pairs of CPUs corresponding to adjacent domains – even if the slabs are one atomic layer thick. Full duplex communication patterns are implemented such that all inter-processor communications can be performed in 2 steps [50].

(2) *Core Algorithms and Memory Requirements.* In the strain computation, the positions of the atoms are computed to minimize the total elastic strain energy. The total elastic energy in the VFF approach has only one, global minimum, and its functional form in atomic coordinates is quartic. The conjugate gradient minimization algorithm in this case is well-behaved and stable. Figure 3 shows the energy convergence behavior in a typical simulation of an InAs/GaAs quantum dot with a total of around 64 million of atoms. The total elastic energy operator is never stored in its matrix form, but the interatomic couplings are computed on the fly. Therefore the only data structures allocated in this phase are the vectors necessary for the conjugate gradient. The implementation used in NEMO 3-D requires six vectors, each of the total size of $3 \times$ number of atoms (to store atomic coordinates, gradients, and intermediate data), however all those vectors are divided into slabs and distributed among CPUs as discussed above. The final atom position vectors are by default stored on all the CPU for some technical output details.

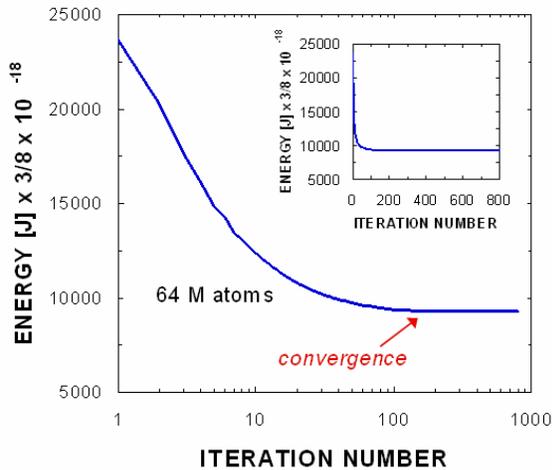

**Figure 3**. Elastic energy convergence profile in a typical simulation of an InAs/GaAs quantum dot with a total 64 million of atoms (inset – linear scale).

The electronic computation involves a very large eigenvector computation (matrices of order of hundreds of millions or even billion). The algorithms/solvers available in NEMO 3-D include the PARPACK library [66], a custom implementation of the Lanczos method, Block Lanczos method, the spectrum folding method [99] and the Tracemin method [86]. The research group is also exploring implementations of Lanczos with deflation method.

The Lanczos algorithm employed here is not restarted, and the Lanczos vectors are not reorthogonalized. Moreover, the spectrum of the matrix has a gap, which lies in the interior of the spectrum. Typically, a small set of eigenvalues is sought, immediately above and below the gap. The corresponding eigenstates are electron and hole wave functions, assuming effectively nonzero values only inside and in the immediate vicinity of the quantum dot. Also, in the absence of the external magnetic field the eigenvalues are repeated, which reflects the spin degeneracy of electronic states. The advantage of Lanczos algorithm is that it is fast, while the disadvantage is that it does not find the multiplicity and can potentially miss eigenvalues. Some comparisons have shown that the Lanczos method is faster by a factor of 40 for the NEMO 3-D matrix than PARPACK. Block Lanczos with block size $p$ finds $p$ degenerate eigenvalues relatively fast compared to PARPACK and Tracemin, however a potential instability exists as well. The Tracemin algorithm finds the correct spectrum of degenerate eigenvalues, but is slower than Lanczos. PARPACK has been found to be less reliable for this problem, taking more time than Tracemin and missing some of the eigenvalues and their multiplicity. Tables II and III give a comparison of Lanczos, Block Lanczos, PARPACK and Tracemin with the number of eigenvalues searched was kept constant. The majority of the memory allocated in the electronic calculation in Lanczos is taken up by the Hamiltonian matrix. This matrix is very large, but typically very sparse; this property is explicitly accounted for in the memory allocation scheme. All matrix entries are, in general, complex, and are stored in single precision. The code has an option to not store the Hamiltonian matrix, but to recompute it, each time it needs to be applied to a vector. In the Lanczos method, this is required once in each iteration. The PARPACK and Tracemin algorithms require the allocation of a significant number of vectors as a workspace, which is comparable to or larger than the Hamiltonian matrix. This additional memory need may require a matrix recompute for memory savings on memory-poor platforms like an IBM BlueGene.

Figure 4 shows the memory requirements for the dominant phase of the code (electronic structure calculations). It shows how the number of atoms that can be treated grows as a function of the number of CPUs, for a fixed amount of memory per CPU. The number of atoms can be intuitively characterized by the length of one side of a cube that would contain that

TABLE II

Performance comparison of different eigenvalue solvers on 32 processors of Purdue University Linux cluster (Xeon x86-64 Dual Core 2.33GHz). Simulation was performed on an InAs QD structure with 268800 atoms. Time (in hours), Relative time, Number of matrix-vector products (#MVP), Relative matrix-vector products, Memory (in GB) and number of correct eigenvalues and their multiplicity (#Eig(mul)) for Lanczos, Block Lanczos with block size 2 (BLanczos2), PARPACK, Tracemin with Quadratic mapping(QTracemin) and Tracemin with Chebyshev polynomial mapping(CTracemin).

| ALGORITHM | TIME (HRS.) | RELATIVE TIME | #MVP ($\times 1000$) | RELATIVE MVP | MEMORY (GB) | #EIG.(MUL) |
|---|---|---|---|---|---|---|
| Lanczos | 0.428 | 1.0 | 10.9 | 1.0 | 2.64 | 20(1) |
| BLanczos2 | 1.385 | 3.2 | 11.8 | 1.1 | 2.77 | 8(2) |
| PARPACK | 18.04 | 42.2 | 59.3 | 5.4 | 2.64 | 8(2),4(1) |
| QTracemin | 15.71 | 36.7 | 317.0 | 29.1 | 2.77 | 10(2) |
| CTracemin | 13.70 | 32.1 | 528.8 | 48.5 | 2.64 | 10(2) |

TABLE III

List of spectrum between 1.0~1.3 eV and the number of multiplicities obtained from different solvers. *Number* of searched eigenvalues was kept constant for these methods.

| EIGENVALUES | LANCZOS | BLANCZOS2 | PARPACK | QTRACEMIN | CTRACEMIN |
|---|---|---|---|---|---|
| 1.0361 | 1 | - | - | 2 | 2 |
| 1.0969 | 1 | 2 | - | 2 | 2 |
| 1.0976 | 1 | 2 | 1 | 2 | 2 |
| 1.1624 | 1 | 2 | 2 | 2 | 2 |
| 1.1645 | 1 | 2 | 2 | 2 | 2 |
| 1.1748 | 1 | 2 | 2 | 2 | 2 |
| 1.2304 | 1 | 2 | 2 | 2 | 2 |
| 1.2312 | 1 | 2 | 2 | 2 | 2 |
| 1.2445 | 1 | 2 | 2 | 2 | 2 |
| 1.2448 | 1 | - | 2 | 2 | 2 |
| 1.2975 | 1 | - | 2 | - | - |

many atoms. This length is shown in Figure 4, on the vertical axis on the right side of each plot. This figure shows that the number of atoms that can be treated in NEMO 3-D continues to grow for larger CPU counts. The strain calculations have so far never been memory limited. NEMO 3-D is typically size limited in the electronic structure calculation.

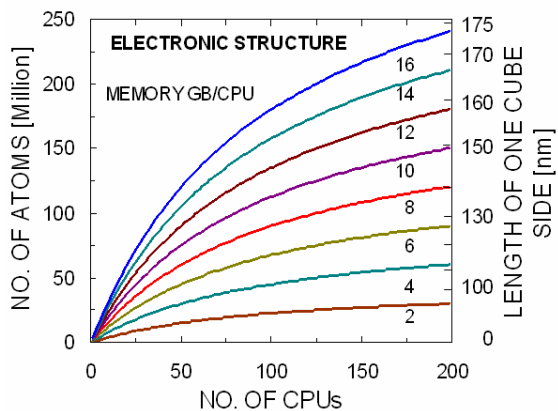

**Figure 4**. Number of atoms that can be treated, as a function of the number of CPUs for different amounts of memory per CPU for the electronic structure calculation. The vertical axis on the right side of each plot gives the equivalent length in nm of one side of the cube that would contain the given number of atoms.

(3) *Optimization in NEMO 3-D*. In running a scientific application that requires massive computation power, we have to consider various issues that may occur, mainly due to limited resource in a computer: Too small memory per core can limit the size of the problem and unnecessary loops in the code consumes additional time for calculation. It is crucial to design an application in a way to maximize floating operations per second and avoid inefficient loops. In NEMO-3D, several optimization ideas are implemented and those are introduced in the following sections.

(a) *Vectorization*. Vectorization is a hardware dependent optimization scheme that converts multiple single scalar operations to single vector operation. The concept is shown in Figure 5. It is commonly used in graphic processors and supercomputers (e.g. Cray X1E machines) where massive computation load and fast processing is needed. Even recent processors in desktop computers, support similar parallel data processing scheme. The most common technique to support parallelism is Single Instruction, Multiple Data (SIMD) algorithm. It was Intel who first developed instruction sets known as Streaming SIMD Extensions, or SSE, to support in their Pentium III processors in 1999 [104]. Nowadays, AMD, Transmeta and Via also support SSE features and new enhancements are developed continuously (as of Oct. 2007, SSE5 is the latest version). A couple of single and double precision arithmetic can be carried out simultaneously resulting in fast computation. Therefore, it is possible to make use of SSE scheme in scientific applications with heavy complex number calculations. In NEMO-3D, complex multiplication and addition occurs frequently in matrix-matrix multiply routine. To this certain application, major improvement was achieved in real-complex multiplies. Figure 6 shows the speed improvement observed in NEMO-3D by replacing SSE instructions to real-complex multiplication.

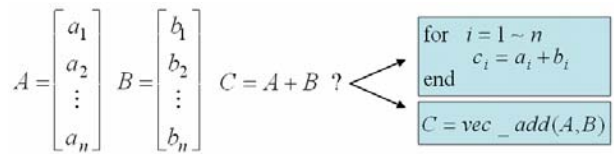

**Figure 5**. The conceptual diagram of vectorization. In vectorized CPU, it is capable of $n$ simultaneous operations in single CPU cycle.

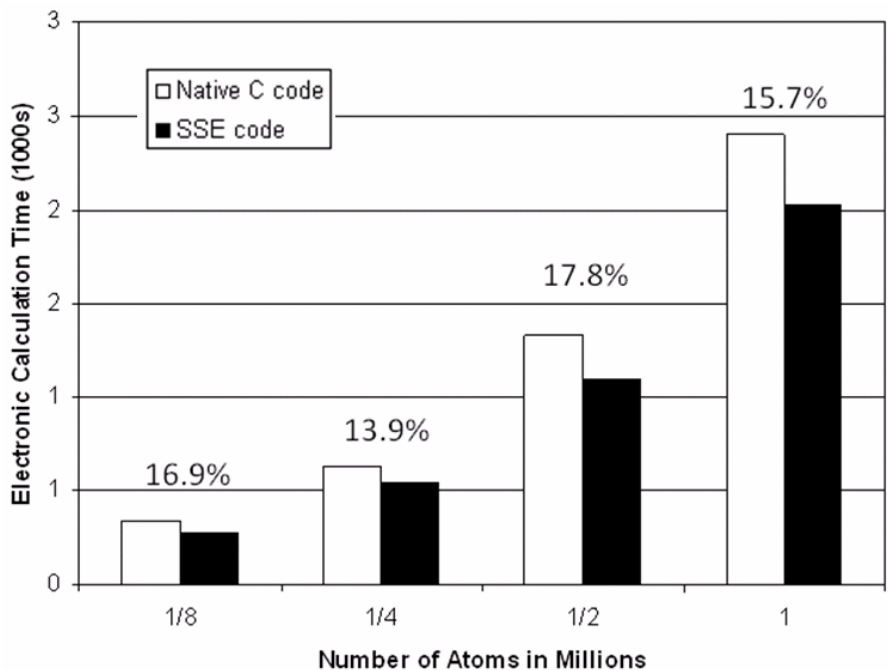

**Figure 6**. Comparison of electronic calculation time between SSE optimized code and native C code. Simulated on a single node of Xeon x86-64 Dual Core 2.33GHz CPU computers.

(b) *Matrix-Matrix Multiplier and BLAS*. The Basic Linear Algebra Subprograms, or BLAS, are standardized interface for performing basic matrix-vector and matrix-matrix multiplication. The BLAS package is widely used in high-performance computing and it has been optimized to maximize the number of floating point operations for specific CPUs. For example, Intel develops its own BLAS package in the Math Kernel Library (MKL BLAS) highly optimized to their processors. Compared to native C code with double nested loops, benefits can be made from BLAS, especially with matrix-matrix multiplication. From the experiment shown in Figure 7, highly-optimized BLAS Matrix-matrix multiply instruction, or ZGEMM, is capable of utilizing the CPU to perform more floating point operations per second, reducing the total calculation time. Even for the block sizes $N = 10$, $N =$ 20 corresponding to $sp^3d^5s^*$ bands significant improvement can be seen by performing block-wise operations. The data in Figure 7 indicates an excellent incentive for the Block Lanczos and the Tracemin algorithms that perform multiple matrix-vector multiplies for the same matrix to be blocked. For example, at $N = 10$ a single vector multiply can be performed at about 1.5 GFlops while 8 multiplies can be performed at a rate of 3.6 GFlops. With the increase in relative performance for increased block size the required total CPU time increased sublinearly. Subsequent NEMO 3-D development for general 3-D spatial structures will utilize the ZGEMM multiply by arranging the data structures such that no copy is needed.

(c) *Explicit Construction of Hamiltonian in Recompute Mode*. The recompute mode enables NEMO-3D to run on limited memory computers by

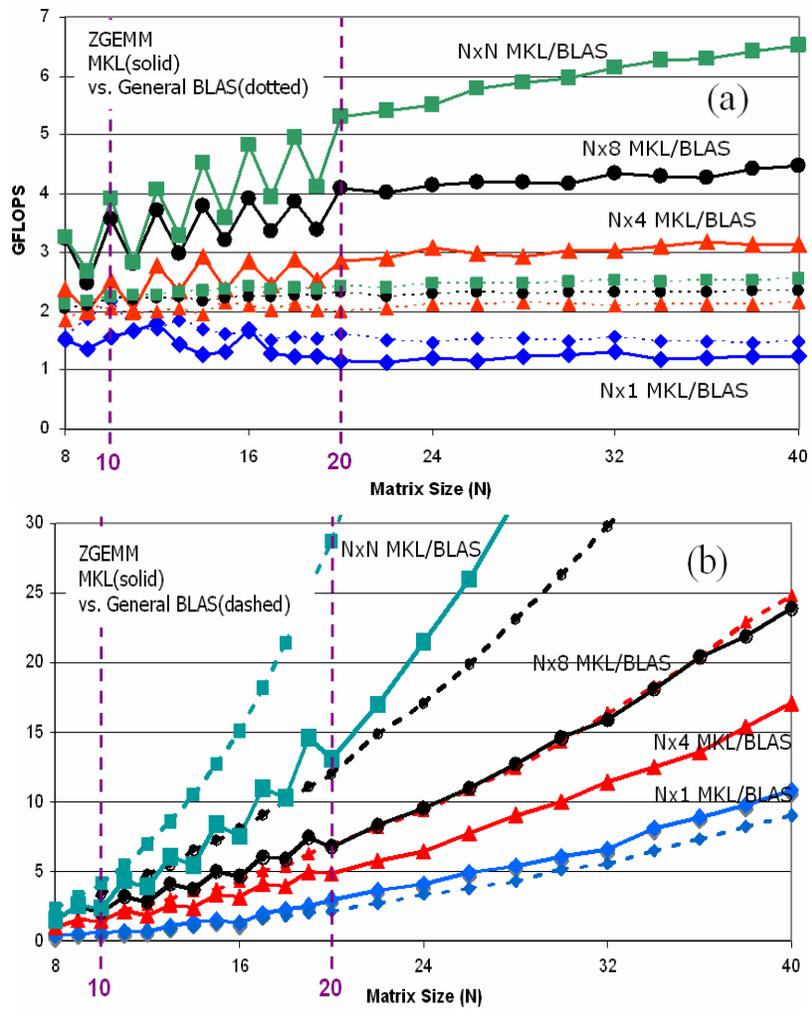

**Figure 7**. (a) Performance plots of ZGEMM (Y=AX) included in different BLAS libraries. GFLOPS ($10^9$ Floating Operations/second) measures of ZGEMM from MKL/BLAS (solid line) and general BLAS/LAPACK library (open markers) are plotted with varying size of $A(N \times N)$ and column size of $X(N \times M)$. Simulated on a single node of Xeon x86-64 Dual Core 2.33 GHz CPU computer. (b) Total compute time of data in (a).

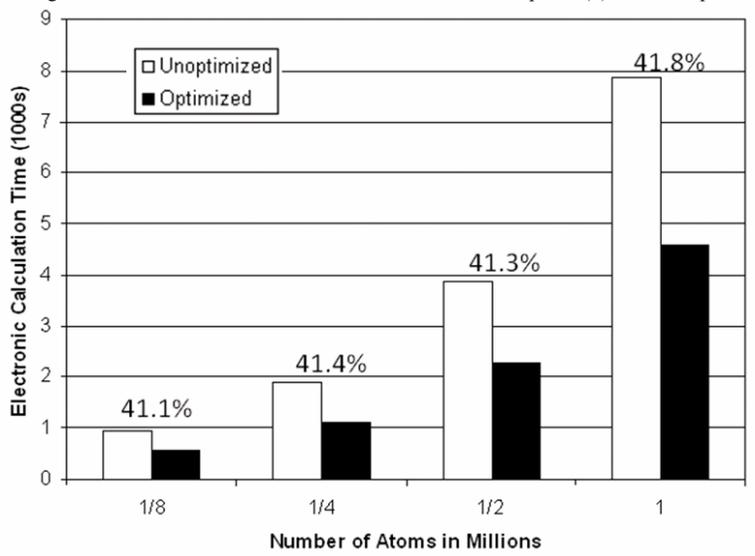

**Figure 8**. The electronic calculation time comparison of optimized/unoptimized Hamiltonian construction in recompute mode. Simulated on 4 nodes of Xeon x86-64 Dual Core 2.33GHz CPU computers.

TABLE IV

Specifications for the HPC platforms used in the performance comparisons.

| PLATFORM | TYPE | CPU | # OF CORES | MEMORY /CORE | INTERCONNECT | TOP 500 JUNE 2007 | LOCATION |
|---|---|---|---|---|---|---|---|
| ORNL/Jaguar | Cray XT3/4 | Opteron x86-64 2.6GHz | 23,016 | 2GB | Native | #2 | ORNL |
| RPI/BGL | BlueGene/L | PowerPC 440 0.7 GHz | 32,768 | 256MB | Native | #7 | RPI |
| IUPU/Big Red | IBM JS21 | PowerPC 970 2.5 GHz | 3,072 | 2GB | Myrinet | #8 | IUPU |
| PSC/XT3 | Cray XT3 | Opteron x86-64 2.6GHz | 4,136 | 1GB | Native | #30 | PSC |
| PU/Xeon D | Linux Cluster | Xeon x86-64 Dual Core 2.33GHz | 672 | 2GB/4GB | Gigabit Ethernet | #46 | RCAC Purdue |

eliminating storage of the Hamiltonian altogether and recomputing the matrix elements as they are needed. However, since the construction of the Hamiltonian consumes significant time, reducing the number of calculations in the construction of a matrix element enhances the performance. In cases where no external magnetic field is present, duplicate calculations due to the spin degeneracy can be avoided. Also, since the orbital interactions are known, unnecessary loops can be avoided, and non-zero elements may be explicitly evaluated. The doubly nested switch statements at the core of the orbital-orbital interaction loops have been replaced by customized expressions for the matrix elements for specific tight binding orbital arrangements such as $sp^3s^*$ and $sp^3d^5s^*$. Simulation result indicates that the electronic calculation time is reduced up to 40% (Figure 8). This customization increases computational performance but reduces the algorithmic generality.

(4) *Scaling*. Out of the two phases of NEMO 3-D, the strain calculation is algorithmically and computationally less challenging than the Lanczos diagonalization of the Hamiltonian matrix.

To investigate the performance of NEMO 3-D package, computation was performed in a single dome shaped InAs quantum dot nanostructure embedded in a GaAs barrier material as shown in Figure 2. The HPC platform used in the performance studies are shown in Table IV. These include a Linux clusters at the Rosen Center for Advanced Computing (RCAC) at Purdue with Intel processors (dual core Woodcrest). The other five platforms are a BlueGene at the Rensselaer Polytechnic Institute (RPI), the Cray XT3 at the Pittsburgh Supercomputing Center (PSC), the Cray XT3/4 at ORNL, JS21 at Indiana University, and a Woodcrest machine at NCSA. Table IV provides the relevant machine details. These platforms have proprietary interconnects, that are higher performance than Gigabit Ethernet (GigE) for the three Linux clusters at Purdue. In the following, the terms processors and cores are used interchangeably.

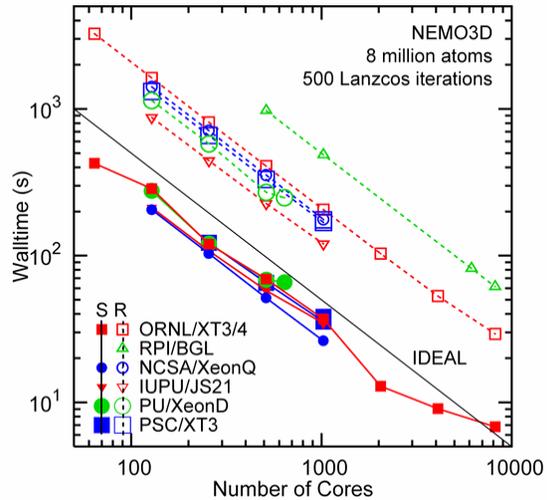

Figure 9. Strong scaling of a constant problem size (8 million atoms) on 6 different HPC platforms. Solid / dashed lines correspond to a stored / recomputed Hamiltonian matrix. The largest number of cores available were 8,192 on Cray XT3/4 and IBM BlueGene.

Figure 9 shows the performance of NEMO 3-D for each of the architectures. The wall clock times for 500 iterations of the Lanczos method for the electronic structure phase are shown as a function of the number of cores. The benchmark problem includes eight million atoms. Figure 9 shows that the PU/Woodcrest cluster is close to the performance of the Cray XT3 for lower core counts, while the XT3 performs better for higher core counts, due to its faster interconnect. The BlueGene's slower performance is consistent with its lower clock speed, while the scalability reflects its efficient interconnect.

Recomputing the Hamiltonian causes a performance reduction of about a factor of 4−6. Since the IBM BlueGene L is memory-poor, we can operate NEMO 3-D only in the Hamiltonian recomputed mode. Since the IBM BlueGene runs about a factor of 4× slower than the other HPC platforms one can see about a factor of 16 × better performance in Cray XT3/4 since it runs fast and has enough memory.

In addition to the performance for the benchmark cases end-to-end runs on the PU/Woodcrest cluster are carried out next (Figure 10). This involves iterating to convergence and computing the eigenstates in the desired range (4 conduction band and 4 valence band states). For each problem size, measured in millions of atoms, the end-to-end cases were run to completion, for one choice of number of cores.

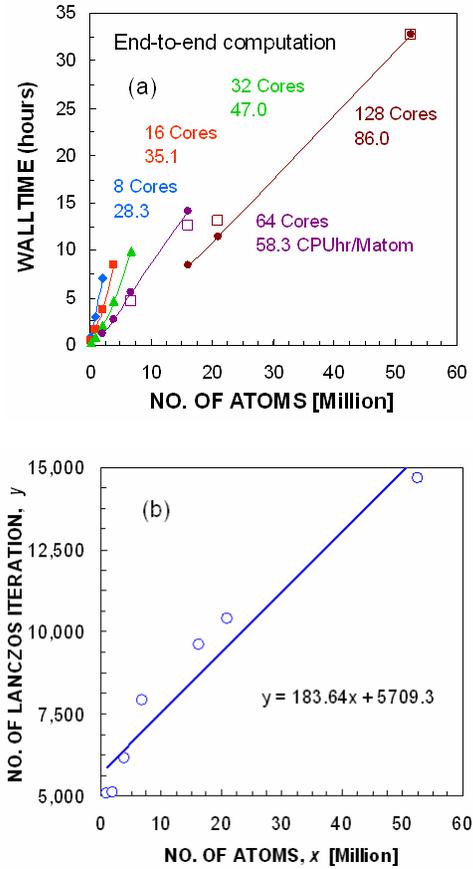

**Figure 10**. (a) Wall clock time vs. number of atoms for end-to-end computations of the electronic structure of a quantum dot, for various numbers of cores on the PU/Woodcrest cluster. Listed next to the number of cores are the CPU hours/Million of atoms needed in the simulation. (b) No. of Lanczos iteration vs. number of atoms for one choice of number of cores.

The numerical experiment is designed to demonstrate NEMO 3-D's ability to extract targeted interior eigenvalues and vectors out of virtually identical systems of increasing size. A single dome shaped InAs quantum dot embedded in GaAs is considered. The GaAs buffer is increased in size to increase the dimension of the system while not affecting confined states in the QD. It is verified [4] that the eigenvectors retain the expected symmetry of the nanostructure.

*(D) Visualization*

The simulation data of NEMO 3-D contains multivariate wave functions and strain profiles of the device structure. For effective 3-D visualizations of these results, a hardware-accelerated direct volume rendering system [80] has been developed, which is combined with a graphical user interface based on *Rappture*. Rappture is a toolkit that supports and enables the rapid development of *graphical user interfaces* (GUIs) for applications, which is developed by Network for Computational Nanotechnology at Purdue University. Two approaches can be followed: (1) The legacy application is not modified at all and a *wrapper script* translates Rappture I/O to the legacy code. (2) Rappture is integrated into the source code to handle all I/O. The first step is to declare the parameters associated with one's tool by describing Rappture objects in the Extensible Markup Language (XML). Rappture reads the XML description for a tool and generates the GUI automatically. The second step is that the user interacts with the GUI, entering values, and eventually presses the Simulate button. At that point, Rappture substitutes the current value for each input parameter into the XML description, and launches the simulator with this XML description as the driver file. The third step shows that, using parser calls within the source code, the simulator gets access to these input values. Rappture has parser bindings for a variety of programming languages, including C/C++, Fortran, Python, and MATLAB. And finally, the simulator reads the inputs, computes the outputs, and sends the results through run file back to the GUI for the user to explore. The visualization system uses data set with *OPEN-DX* format that are directly generated from NEMO 3-D. *OPEN-DX* is a package of open source visualization software based on IBM's Visualization Data Explorer. Figure 11 shows the wave functions of electron on the first 4 eigenstates in conduction band of quantum dot which has 268,800 atoms in the electronic domain.

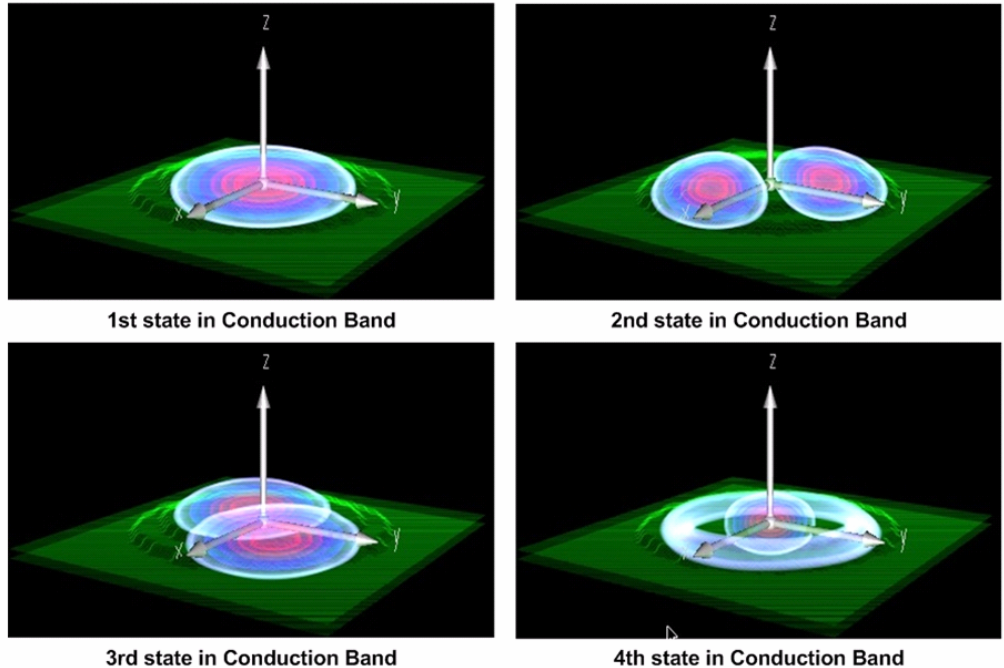

**Figure 11**. Wave function profiles of first 4 electron eigenstates in the conduction band. Green color shows active InAs region where confinement takes place.

*(E) Release and Deployment of NEMO 3-D Package*

NEMO 3-D was developed on Linux clusters at the Jet Propulsion Lab (JPL) and was released with an open source license in 2003. The originally released source is hosted at http://www.openchannelfoundation.org web site. As NEMO 3-D is undergoing further developments by the NCN we are planning future releases of the NEMO 3-D source through http://www.nanoHUB.org. NEMO 3-D has been ported to different high performance computing (HPC) platforms such as the NSF's TeraGrid (the Itanium2 Linux cluster at NCSA), Pittsburgh's Alpha cluster, Cary XT3, SGI Altix, IBM p690, and various Linux clusters at Purdue University and JPL.

The NEMO 3-D project is now part of a wider initiative, the NSF Network for Computational Nanotechnology (NCN). The main goal of this initiative is to support the National Nanotechnology Initiative through research, simulation tools, and education and outreach. Deployment of these services to the science and engineering community is carried out via web-based services, accessible through the nanoHUB portal http://www.nanoHUB.org. The educational outreach of NCN is realized by enabling access to multimedia tutorials, which demonstrate state-of-the-art nanodevice modeling techniques, and by providing space for relevant debates and scientific events. The second purpose of NCN is to provide a comprehensive suite of nano simulation tools, which include electronic structure and transport simulators of molecular, biological, nanomechanical and nanoelectronic systems. Access to these tools is granted to users via the web browsers, without the necessity of any local installation by the remote users. The definition of specific sample layout and parameters is done using a dedicated Graphical User Interface (GUI) in the remote desktop (VNC) technology. The necessary computational resources are further assigned to the simulation dynamically by the web-enabled middleware, which automatically allocates the necessary amount of CPU time and memory. The end user, therefore, has access not only to the code, a user interface, and the computational resources necessary to run it but also to the scientific and engineering community responsible for its maintenance. The nanoHUB is currently considered one of the leaders in science gateways and cyber infrastructure.

The process of web-based deployment of these tools is depicted in Figure 12. A user visits the www.nanohub.org site and finds a link to a tool. Clicking on that link will cause our middleware to create a virtual machine running on some available CPU. This virtual machine gives the user his/her own private file system. The middleware starts an application and exports its image over the Web to the user's browser. The application looks like an Applet running in the browser. The user can click and interact with the application in real time taking advantage of high-performance distributed computing power available on local clusters at Purdue University and on the NSF TeraGrid or the open science grid.

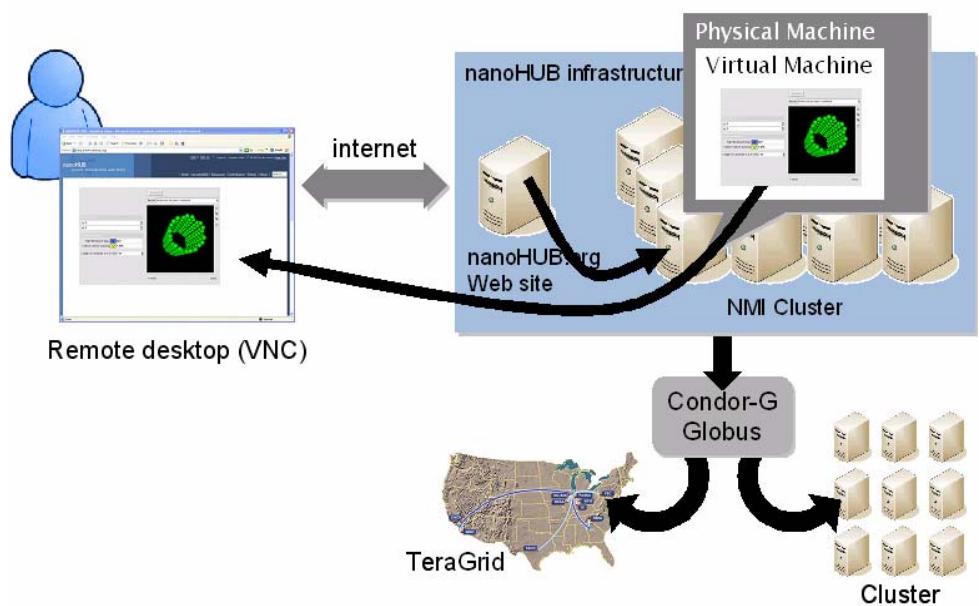

**Figure 12**. Deployment of the NCN nanotechnology tools on http://www.nanoHUB.org: Remote access to simulators and compute power.

Recently, a prototype graphical user interface (GUI) based on the *Rappture* package (http://www.rappture.org) is incorporated within the NEMO 3-D package and a web-based online *interactive* version (Quantum Dot Lab) for educational purposes is freely available on www.nanohub.org [38]. The currently deployed NEMO 3-D educational version is restricted to a single *s* orbital basis (single band effective mass) model and runs in seconds. Users can generate and freely rotate 3-D wavefunctions interactively powered by a remote visualization service. Quantum Dot Lab was deployed in November 2005 and has been a popular tool used by 1,541 users who ran 12,616 simulations up to August 2008. Monthly and annualized users and simulation numbers are shown in Figure 13.

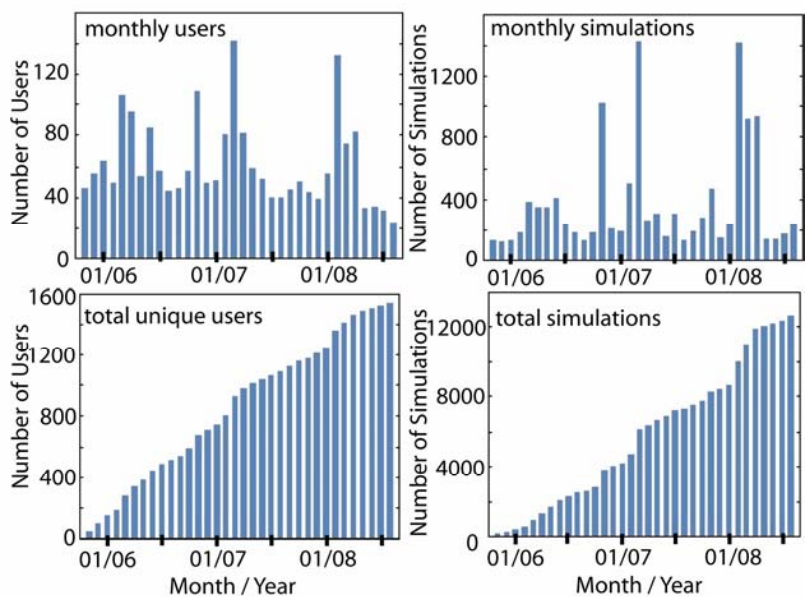

**Figure 13**. (first row) Number of *monthly* users who have run at least one simulation and number of *monthly* simulation runs executed by nanoHUB users. (second row) Number of *total* users who have run at least one simulation and *total* simulation runs executed by nanoHUB users.

The complete NEMO 3-D package is available to selected members of the NCN community through the use of a nanoHUB workspace. A nanoHUB workspace presents a complete Linux workstation to the user within the context of a web browser. The workstation persists beyond the browser lifetime enabling to user to perform long duration simulations without requiring their constant attention. As shown in this paper the computational resources required to perform device scale simulations are considerable and beyond the reach of many researchers. With this requirement in mind NCN has joined forces with Teragrid [94] and the Open Science Grid [73] to seamlessly provide the necessary backend computational capacity to do computationally intensive computing. Computational resources necessary for large scale parallel computing are linked to nanoHUB through the Teragrid *Science Gateways* program. Access to a Teragrid allocation is provided for members of the NCN community. Development of a more comprehensive NEMO 3-D user interface continues. The more comprehensive interface will provide access to a broader audience and encourage the continued growth of the nanoHUB user base.

## V   SIMULATION RESULTS

### (A)   Strain and Piezoelectricity in InAs/GaAs Single QDs

The dome shaped InAs QDs that are studied first in this work are embedded in a GaAs barrier material (schematic shown in Figure 2) and have diameter and height of 11.3 nm and 5.65 nm respectively, and are positioned on a 0.6-nm-thick wetting layer [6][60]. The simulation of strain is carried out in the larger computational box (width $D_{strain}$ and height $H$), while the electronic structure computation is usually restricted to the smaller domain (width $D_{elec}$ and height $H_{elec}$). All the strain simulations in this category fix the atom positions on the bottom plane to the GaAs lattice constant, assume periodic boundary conditions in the lateral dimensions, and open boundary conditions on the top surface. The inner electronic box assumes closed boundary conditions with passivated dangling bonds [61]. The strain domain contains ~3 M atoms while the electronic structure domain contains ~0.3M atoms.

*Impact of Strain.* Strain modifies the effective confinement volume in the device, distorts the atom bonds in length and angles, and hence modulates the local Bandstructure and the confined states. Figure 14 shows the diagonal (biaxial) components of strain distribution along the [001] direction in the quantum dot (cut through the center of the dot). There are two salient features in this plot: (a) The atomistic strain is long-ranged and penetrates deep into both the substrate and the cap layers, and (b) all the components of biaxial stress have a non-zero slope inside the quantum dot region. The presence of the gradient in the trace of the hydrostatic strain introduces unequal stress in the zincblende lattice structure along the depth, breaks the equivalence of the [110] and [1$\underline{1}$0] directions, and finally breaks the degeneracy of the first excited electronic state (the so-called $P$ level). Figure 15a shows the wavefunction distribution for the first 8 (eight) conduction band electronic states within the device region for the dot (in a 2-D projection). Note the optical anisotropy and non-degeneracy in the first excited ($P$) energy level. The first $P$ state is oriented along the [110] direction and the second $P$ state along the [1$\underline{1}$0] direction. The individual energy spectrum is also depicted in this figure which reveals the value of the $P$ level splitting/non-degeneracy (defined as $E_{1\underline{1}0} - E_{110}$) to be about 5.73 meV.

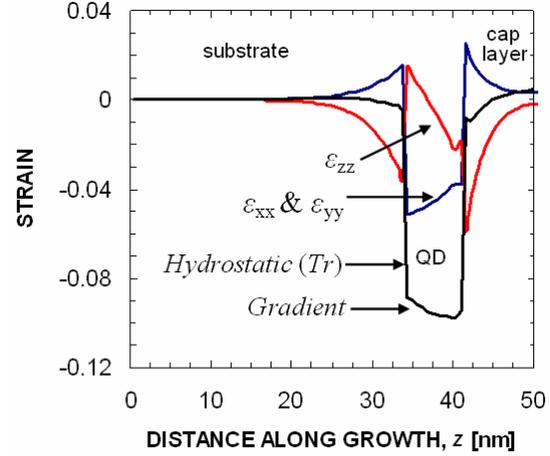

**Figure 14**. Atomistic *diagonal* strain profile along the [001], $z$ direction. Dome shaped dot with Diameter, $d$ = 11.3 nm and Height, $h$ = 5.65 nm. Strain is seen to penetrate deep inside the substrate and the cap layer. Also, noticeable is the gradient in the trace of the hydrostatic strain curve ($Tr$) inside the dot region that results in optical polarization anisotropy and non-degeneracy in the electronic conduction band $P$. Atomistic strain thus lowers the symmetry of the dot.

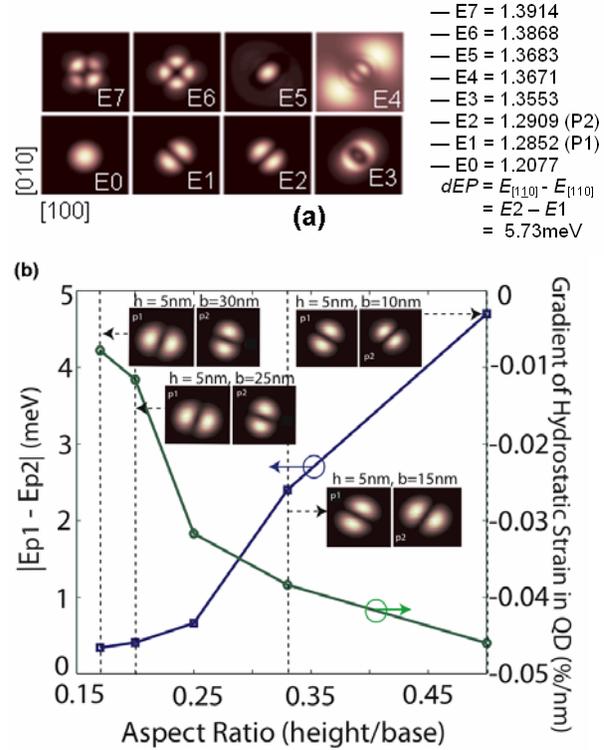

**Figure 15.** (a) Conduction band wavefunctions and spectra (eV) for first eight energy levels in the Dome shape quantum dot structure. Atomistic strain is included in the calculation. Note the optical anisotropy and non-degeneracy in the *P* energy level. The first state is oriented along [110] direction and the second state along [1$\bar{1}$0] direction. (b) gradient in the hydrostatic strain along the [001] direction through the center of the dot and the resulting non-degeneracy and optical anisotropy in the *P* level as a function of the dot aspect ratio.

As explained in Ref. [6], the shape-symmetry of a quantum dot is lowered due mainly to three reasons, all originating from the fundamental atomistic nature of the underlying crystal: (1) The *interface* between the dot material (InAs) and the barrier material (GaAs), even with a common anion (As atom), is not a reflection plane and hence anisotropic with respect to the anion. The direct neighbors above the anion plane (In atoms) that align in the [$\underline{1}$10] direction are chemically different from the neighbors under the anion plane (Ga atoms) that align in the [110] direction. This creates a short-range interfacial potential. It is important to note that these atomistic interfacial potentials originating from different facets do not necessarily compensate each other in dots where the base is larger than the top (for example, pyramid, lens, truncated pyramid). (2) *Atomistic strain and relaxations* (originating from the atomic size difference between Ga and In atoms) results in a propagation of the interfacial potential further into the dot material and thus amplifies the magnitude of the asymmetry. This component is not captured if the relaxation is performed using classic harmonic continuum-elasticity approach. Noticeable is the fact that, symmetry breaking due to atomistic relaxations can even be observed in dots where the base is equal to the top (for example, box, disk); however, the effect is magnified in dots of typical shape, where the base is larger than the top (for example, pyramid, lens, truncated pyramid) due to the presence of a gradient in the magnitude of the strain tensor between top and bottom as already explained in Figure 15a. In order to further characterize this effect, we have simulated dome-shaped dots with varying base diameters (from 10 to 30 nm) keeping the dot height constant (at 5 nm). Figure 15b shows the gradient in the hydrostatic strain and the resulting non-degeneracy in the $P$ level as a function of the dot aspect ratio (height/base). Also, shown in the insets are the wavefunctions corresponding to the split $P$ levels in each of these dots. Note that the non-degeneracy and the optical anisotropy diminish as the dot aspect ratio decreases (approaching a disk shape). (3) Finally, a long-ranged *piezoelectric field* develops in these dots in response to the strain-induced displacement field, which is fundamentally anisotropic. We will discuss this effect in detail in a subsequent section.

*Need for a Deep Substrate and a Realistic Cap Layer.* The strength of the NEMO 3-D package lies particularly in its capability of simulating device structures with realistic boundary conditions. Our simulation results based on NEMO 3-D show a significant dependence of the dot states and magnitude of level-splitting on the substrate layer thickness, $s$ (underneath the dot) and the cap layer thickness, $c$ (above the dot). The strain in the QD system therefore penetrates deeply into the substrate and cannot be neglected. Figure 16 shows such observed dependency where $E0$ is the ground state energy and $dEP$ is the magnitude of the level splitting in the $P$ electronic states due to the inclusion of atomistic strain and relaxation. The changes in both these quantities are calculated with respect to the largest $s$ (50 nm) and $c$ (20 nm) respectively in Figures 16a and 16b. The wavefunction orientation was found to remain unchanged irrespective of the substrate depth and cap layer thickness. Figure 16a shows that it is indeed important to include enough of a substrate to capture the long-range strain, while Figure 16b indicates opportunities to tune the eigen energy spectrum with different capping layer thicknesses.

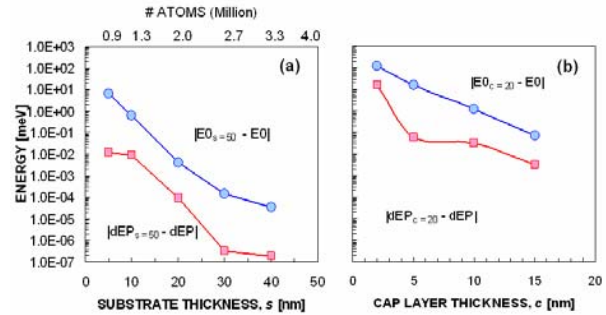

**Figure 16**. (a) Substrate layer thickness dependence of the conduction band minimum and the $P$ level splitting. Other structural parameters remain constant ($h$ = 5.65 nm, $d$ = 11.3 nm, $c$ = 10 nm, and $D$ = 31.3 nm). (b) The impact of cap layer thickness (with substrate, $s$ = 30 nm and other structural parameters remaining the same). Lanczos convergence tolerance = $1 \times 10^{-7}$.

Figure 17 reveals the reason of a strong dependency of the electronic ground state and the magnitude of non-degeneracy in $P$ level on the cap layer thickness. Here the hydrostatic strain profiles for two different cap layer thicknesses (2 nm and 10 nm) are plotted. The $P$ level splitting in a device with 10 nm cap layer is found to be 5.73 meV and that for a 2 nm cap layer was 20.58 meV. The reason of the reduction in the splitting in the 10 nm cap layer device can be attributed mainly to the change in the gradient of hydrostatic strain inside the device region as depicted in Figure 17.

of the $P$ level. Also shown in Figure 21 is the asymmetry in potential profile due to atomistic strain and inequivalence in the piezoelectric potential along [110] and [1$\underline{1}$0] directions at a certain height $z = 1$ nm from the base of the dot.

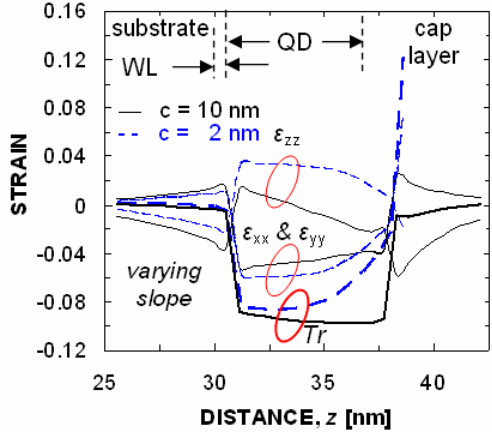

**Figure 17**. The impact of cap layer thickness (with deep substrate, $s = 30$ nm, and $h = 5.65$ nm, $d = 11.3$ nm). Shown is the significant variation of gradient/slope in the strain profile within the quantum dot region. This results in a different splitting in the conduction band P energy level for the two different thicknesses of the cap layer.

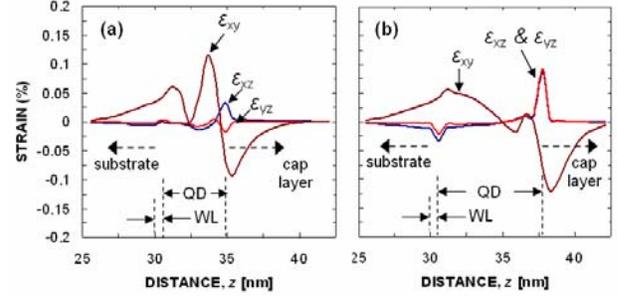

**Figure 18**. Atomistic *off-diagonal* strain profile along the $z$ (vertical) direction which in effect induces polarization in the quantum dot structure. (a) Diameter, $d = 11.3$ nm and Height, $h = 2.8$ nm and (b) Diameter, $d = 11.3$ nm and Height, $h = 5.65$ nm. Note the increase in *off-diagonal* stain in (b).

*Impact of Piezoelectric Fields*. The presence of non-zero off-diagonal strain tensor elements leads to the generation of a piezoelectric field in the quantum dot structure, which is incorporated in the simulations as an external potential by solving the Poisson equation on the zincblende lattice. Figures 18a and 18b show the atomistic off-diagonal strain profiles in dome shaped quantum dots with heights, $h$ of 2.8 nm and 5.65 nm respectively. The off-diagonal strain tensors are higher in the larger diameter dot. The off-diagonal strain tensors are found to be larger in the dome shaped dot. The off-diagonal strain tensors are used to calculate the first-order polarization in the underlying crystal (see Ref. [6] for the governing equations) which gives rise to a piezoelectric charge distribution throughout the device region and then used to calculate the potential by solving the Poisson equation. The relevant parameters for the piezoelectric calculation are taken from Ref. [6]. Experimentally measured polarization constants of GaAs and InAs materials (on unstrained bulk) values of -0.16 C/m$^2$ and -0.045 C/m$^2$ are used. The second order piezoelectric effect [5] is neglected here because of unavailability of reliable relevant polarization constants for an InAs/GaAs quantum dot structures.

The calculated piezoelectric charge and potential surface plots in the *XY* and *XZ* planes are shown in Figures 19 and 20 respectively revealing a pronounced polarization effect induced in the structure. It is found that piezoelectric field alone favors the [1$\underline{1}$0] orientation

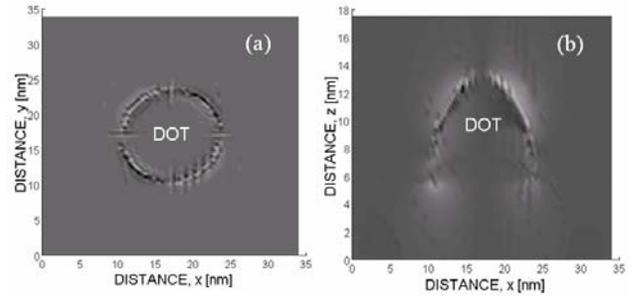

**Figure 19**. Charge surface plot of a dome shape quantum dot (a) in the *XY* plane at $z = 1$ nm from the base of the dot, and (b) in the *XZ* plane at $y = D_{strain}/2$. Charge is induced mainly in the vicinity of the boundary of the quantum dot. ($d = 11.3$ nm and $h = 5.65$ nm).

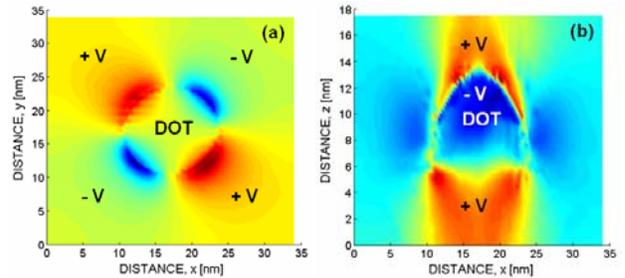

**Figure 20**. Piezoelectric potential surface plot of a dome shape quantum dot (a) in the *XY* plane at $z = 1$ nm from the base of the dot, and (b) in the *XZ* plane at $y = D_{strain}/2$. (c) Potential along [110] and [1$\underline{1}$0] directions at z = 1 nm from the base of the dot. Note the induced polarization in the potential profile and the unequal values of potential along the [110] and [1$\underline{1}$0] directions ($d = 11.3$ nm and $h = 5.65$ nm).

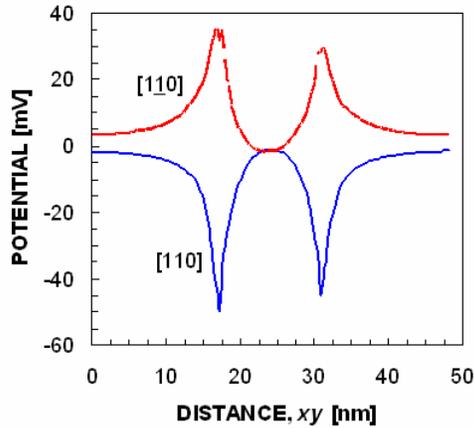

**Figure 21**. Potential along [110] and [1$\bar{1}$0] directions at $z = 1$ nm from the base of the dot. Note the induced polarization in the potential profile and the unequal values of potential along the [110] and [1$\bar{1}$0] directions ($d = 11.3$ nm and $h = 5.65$ nm).

*Study of Varying Sized Dots*. The impact of atomistic strain and piezoelectric field on the ground state energy and magnitude of the $P$ level energy splitting in dome shaped quantum dots with varying diameter $d$ and dot height $h$ is shown in Figures 22 and 23 respectively. The ground state energy for the strained system (without piezoelectricity), $E0$, decreases with an increase in both $d$ and $h$ because of an increase in the effective confinement volume. Figures 22a and 23a also show the change (absolute and relative to strain only) in the ground state energy due to the inclusion of piezoelectric potential in the strained system. The percentage change in the ground state energy is found to be monotonous in nature with an increase in dot diameter while the height dependency shows saturation beyond a certain value. Figures 22b and 23b show the change of three quantities related to the first excited $P$ level namely split due to strain only (circle), split due to strain combined with piezoelectricity (square) and the contribution of the piezoelectric field only (triangle), as a function of diameter $d$ and dot height $h$. The piezoelectric potential introduces a global shift in the energy spectrum, and is found to be strong enough to flip the optical polarization in certain sized quantum dots. In those cases the piezoelectric contribution (triangle) dominates over that resulting from the inclusion of atomistic strain alone in the simulations (circle) as can be seen in dots (see Figure 22b; similar trend has also been found in Ref. [6]) with diameters larger than 7 nm and (see Figure 23b) height more than 3 nm. Figure 24 explains the reason behind this observation. Here the piezoelectric potential profiles in dots with different height $h$ are plotted along the $z$ direction through the dot center. Note the increase in piezoelectric potential with dot height. The stronger piezoelectric potential induced in the larger dot results in the orientational flip in the $P$ level electronic states.

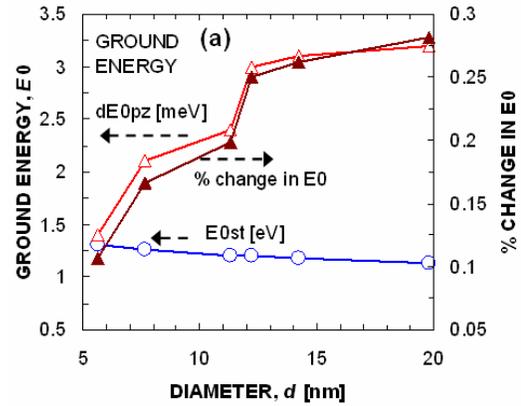

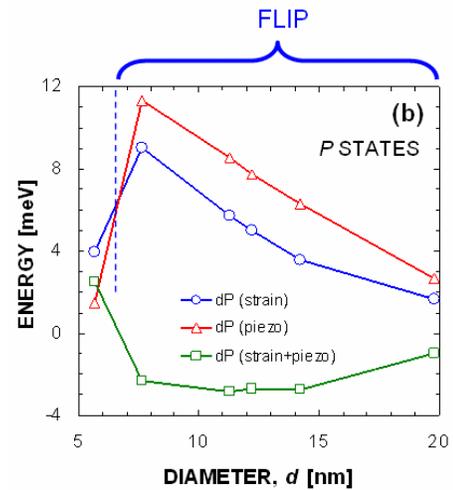

**Figure 22**. Study of electronic structure with the variation of *dot diameter*, $d$ of the dome shaped quantum dot. (a) Conduction band minimum/ground state in a strained system (circle) and change in the conduction band minimum due to induced piezoelectricity (triangle). (b) Split in the $P$ level due to strain only (circle), split in the $P$ level due to strain and piezoelectricity (square), and impact of piezoelectric potential *alone* (triangle) in the system (dot height, $h = 5.65$ nm).

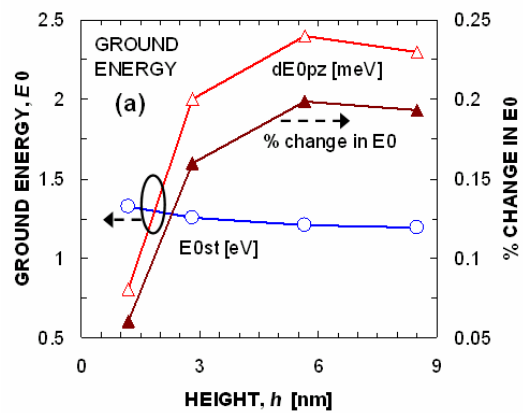

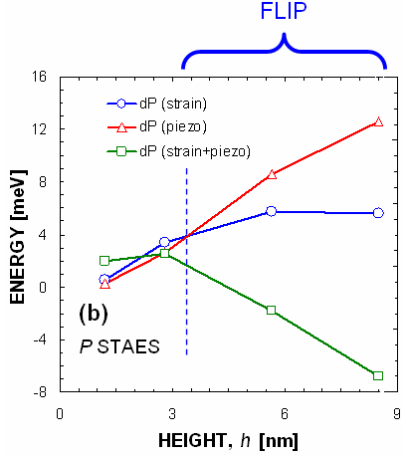

**Figure 23**. Study of electronic structure with the variation of *dot height*, $h$ of the dome shaped quantum dot. (a) Conduction band minimum/ground state in a strained system (circle) and change in the conduction band minimum due to induced piezoelectricity (triangle). (b) Split in the $P$ level due to strain only (circle), split in the $P$ level due to strain and piezoelectricity (square), and impact of piezoelectric potential *alone* (triangle) in the system (dot diameter, $d = 11.3$ nm).

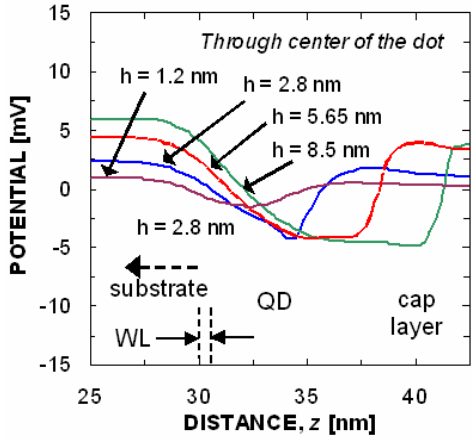

**Figure 24**. Piezoelectric potential in dome shaped quantum dots with $h = 2.8$ nm and $h = 5.65$ nm along the $z$ direction through the center of the dots. Noticeable is the stronger polarization in the larger dot which results in a *flip* in the $P$ level electronic states.

*Piezoelectricity Induced Polarization Flip*. Figure 25 shows the conduction band wavefunctions for the ground and first three excited energy states in the quantum dot structure with diameter of 11.3 nm and height, $h$ of 5.65 nm. In Figure 25a strain and piezoelectricity are *not* included in the calculation. The weak anisotropy in the $P$ level is due to the atomistic interface and material discontinuity. Material discontinuity mildly favors the [110] direction in the dot. In Figure 25b atomistic strain and relaxation is included resulting in a 5.73 meV split in the $P$ energy levels. Strain favors the [110] direction. In Figure 25c piezoelectricity is included on top of strain inducing a split of -2.84 meV in the $P$ energy level. The first $P$ state is oriented along [1$\underline{1}$0] direction and the second state along [110] direction and piezoelectricity alone induces a potential of 8.57 meV. Piezoelectricity thereby has not only introduced a global shift in the energy spectrum but also *flipped* the orientation of the $P$ states [6] in this case.

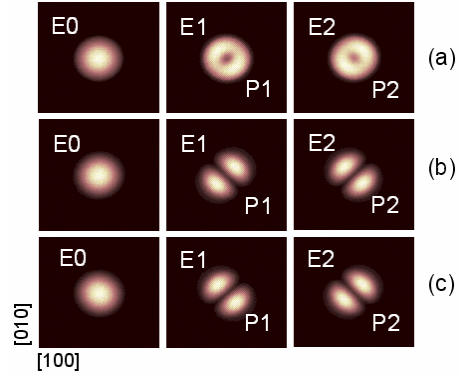

**Figure 25**. Conduction band wavefunctions for first three energy levels in the quantum dot structure with diameter, $d = 11.3$ nm and height, $h = 5.65$ nm (a) without strain and piezoelectricity, $E_{[1\underline{1}0]} - E_{[110]} = 1.69$ meV (b) with atomistic strain, $E_{[1\underline{1}0]} - E_{[110]} = 5.73$ meV and (c) with strain and piezoelectricity, $E_{[1\underline{1}0]} - E_{[110]} = -2.84$ meV. Piezoelectricity *flips* the wavefunctions. An end-to-end computation involved about 4M atoms and needed CPU time of about 8 hours with 16 processors.

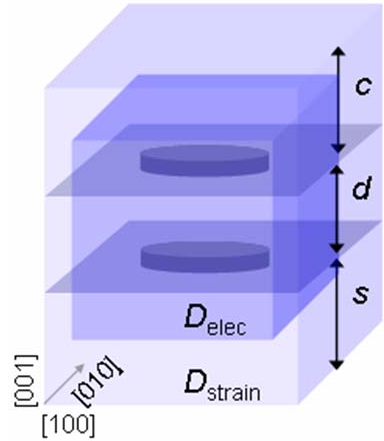

**Figure 26**. Simulated InAs/GaAs double quantum dots with disk/cylindrical shape. The dots are of equal size with radius $r$ of 7nm and height $h$ of 1.5nm. The separation $d$ is varied from 0.5 nm to 8 nm. Two simulation domains have been shown. The strain domain for 8 nm spacing between the dots contained about 6 million atoms.

### (B) Stacked Quantum Dot System

Self-assembled quantum dots can be grown as stacks where the QD distance can be controlled with atomic layer control. This distance determines the interaction of the artificial atomic states to form artificial molecules.

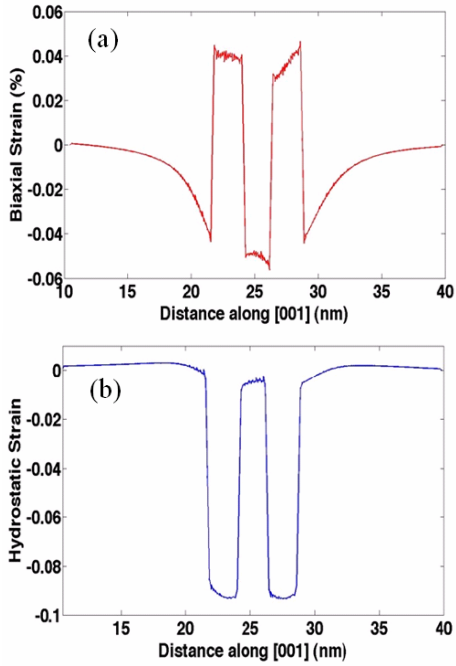

The design of QD stacks becomes complicated since the structures are subject to inhomogeneous, long-range strain and growth imperfections such as non-identical dots and inter-diffused interfaces. Quantum dot stacks consisting of two QD layers are simulated next (see Figure 26). The InAs quantum dots are disk shaped with diameter 7 nm and height 1.5 nm positioned on a 0.6 nm thick wetting layer. The substrate thickness under the first wetting layer is kept constant at 30 nm and the cap layer on top of the topmost dot is kept at 20 nm for all the simulations. The strain simulation domain ($D_{strain}$) contains 8–10 M atoms and the electronic structure domain ($D_{elec}$) contains 0.5–1.1M atoms.

Figure 27 shows both the biaxial and hydrostatic strain profiles along the $z$ direction. As in the single dot, we see a gradient in strain profile within the dot regions which results in strain-induced asymmetry. The hydrostatic component which is responsible for conduction band well is negative within the dot and approximately zero outside the dot and the regions in-between the dots. The biaxial component of strain which have more effect on hole states is positive within dots and negative in-between the dots. The magnitude for both is approximately equal. Figure 28 shows the band edge diagrams as a function of dot separations along the center of the dots in the growth direction [001]. Strain enhances the coupling between the dots. Hydrostatic component of strain makes the conduction band well shallower. Strain effects are more prominent on hole states where biaxial component of strain splits the light hole and heavy hole bands. Within the dot, heavy hole lies above the light hole edge, implying significant band mixing in the confining states. For very small separation like 0.5nm, the well within dots is even shallower than the well in-between the dots. Figure 29 and Figure 30 show the electron and hole state energies respectively as a function of inter-dot separation. In a system without inhomogeneous strain one would expect the identical dots to have degenerate eigenstate energies for large dot separations. Strain breaks the degeneracy even for large separations. As the dot separation is narrowed the dots interact with each other mechanically through the strain field as well as quantum mechanically through wavefunction overlaps. Wave function plots in XZ plane have been shown in Figure 29 for various dot separations. Noticeably, E2 for 4nm separation is a *p* like state while it is *s* like state in 6nm separation. So there is a crossover between *p* to *s* for E2 as we increase separation between the dots. Also, E1 for 2nm separation is confined more in the lower dot than the upper dot. This is caused by strain coupling which promotes confinement of the ground states in the lower dots in coupled quantum dot systems [53]. The electronic states and wavefunctions in a coupled QD system are thus determined through a complicated interplay of strain, QD size, and wavefunction overlap. Only a detailed simulation can reveal that interplay.

**Figure 27**. Atomistic (a) *biaxial* $\{\varepsilon_{zz} - (\varepsilon_{xx}+\varepsilon_{yy})/2\}$ and (b) *hydrostatic* $\{\varepsilon_{xx} + \varepsilon_{yy} + \varepsilon_{zz}\}$ strain profile along the growth [001], $z$ direction. Strain is seen to penetrate deep inside the substrate and the cap layer. Also, noticeable is the gradient in the trace of the hydrostatic strain curve ($Tr$) inside the dot region that results in optical polarization anisotropy and non-degeneracy in the electronic conduction band $P$. Atomistic strain thus lowers the symmetry of the dot.

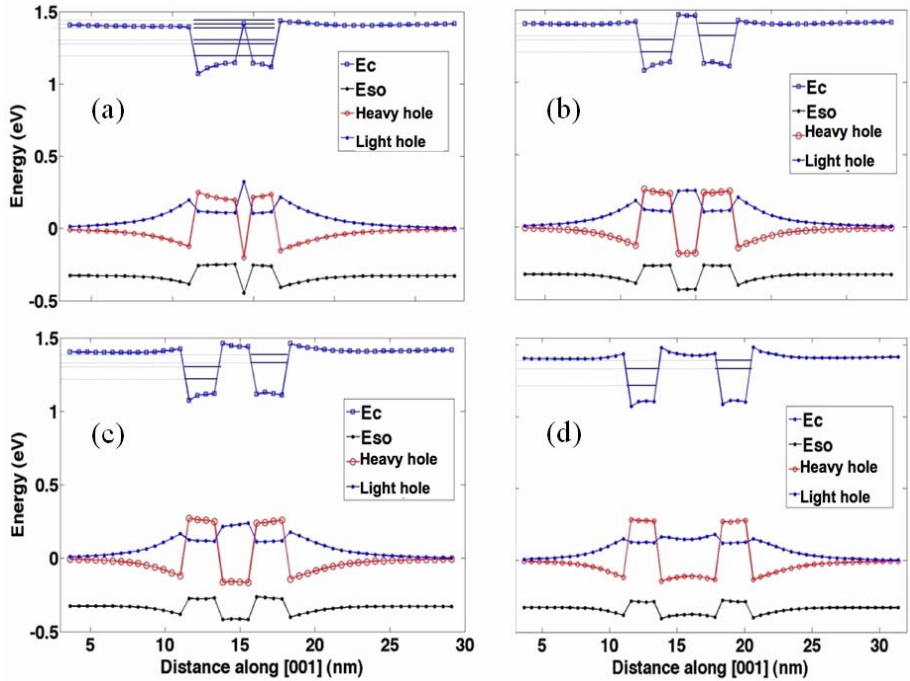

**Figure 28**. Band edge diagrams for double quantum dot systems for several inter-dot spacing: (a) 0.5nm (b) 1nm (c) 2nm and (d) 4nm. Strain makes InAs conduction band potential wells shallower, enhancing the coupling between the dots. Noticeable is the effect on hole wells. Strain splits the light hole and heavy hole bands. Within the dot, heavy hole lies above the light hole edge. As strain coupling decreases, heavy hole well become more and more shallower (see b and d).

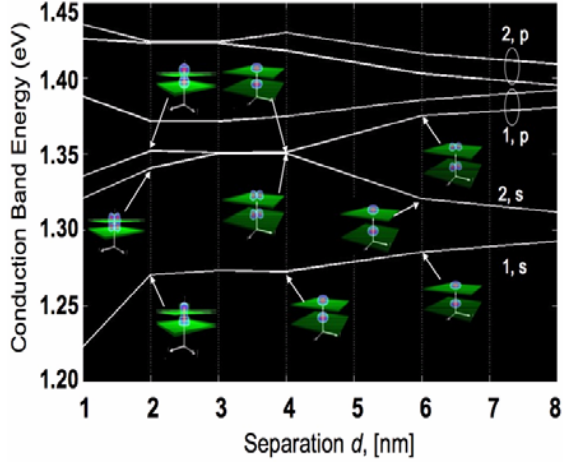

**Figure 29**. Dependence of six lowest electron energy levels on separation distance *d* between the dots. For electron energy levels, the state names are mentioned as *s* or *p* orbital states. Here 1 indicates bonding states whereas 2 indicates anti-bonding states. Wave function plots in XZ plane have been shown for some dot separations. Noticeably, e2 for 4nm separation is a p like state while it is s like state in 6nm separation. So there is a crossover between p to s for e2 as we increase separation between the dots. Also, e1 for 4nm separation is confined in lower dot more than upper dot. This is caused by strain coupling which tries to confine ground states in the lower dots in coupled quantum dot systems.

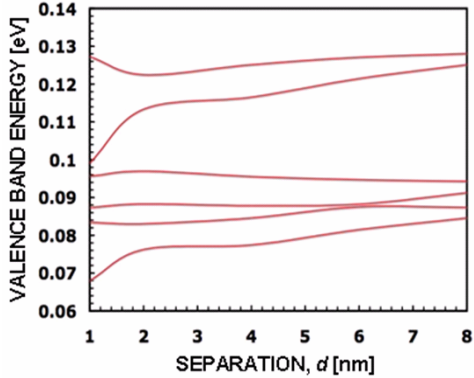

**Figure 30**. Dependence of six lowest hole energy levels on separation distance *d* between the dots.

### (C) Phosphorus (P) Impurity in Silicon

*Physics of P Impurity*. In a substitutional P impurity in Si, the 4 electrons from the outermost shell of P form bonds with the 4 neighboring Si atoms, while the $5^{th}$ electron can ionize to the conduction band at moderate temperatures leaving a positively charged P atom with a coulomb potential screened by the dielectric constant of the host. At low temperatures, this potential can trap an electron, and form an Hydrogen-like system except the six fold degenerate conduction band valleys of Si give rise to a six fold degenerate 1s type ground state. In practice, this six fold degeneracy is lifted by strong coupling between the different valleys caused by deviations of the impurity potential from its coulombic nature in the vicinity of the donor nucleus. If this so called valley-orbit interaction is not taken into account, then the effective mass theory (EMT) predicts a P donor ground state binding energy of -33 meV as opposed to the experimentally measured value of -45.6 meV [87]. The influence of valley-orbit interaction is strongest for the six 1s states, and is negligible for the excited states, which are affected by the bulk properties of the host [15]. The TB model considered here also models the excited states well by its accurate representation of the Si band structure. Hence, we limit our attention here to the effect of valley-orbit interaction on the 1s states.

*Study Approaches*. Theoretical study of donors in Si dates back to the 1950s when Kohn and Luttinger [51] employed symmetry arguments and variational envelope functions based on EMT to predict the nature of the donor spectrum and wave functions with a fair amount of success. Although many theorists who study donor based nano devices still use the Kohn-Luttinger variational envelope functions, recent approaches [65][82][100] have highlighted the need to consider a more extended set of Bloch states than the six valley minima states and to go beyond the basic EMT assumptions for accurate modeling of impurities. For modeling high precision donor electronics, it is very important to model the basic Physics from a consistent set of assumptions, and to obtain very accurate numbers in addition to correct trends. The model presented here serves these purposes well, and can be used conveniently for large-scale device simulation.

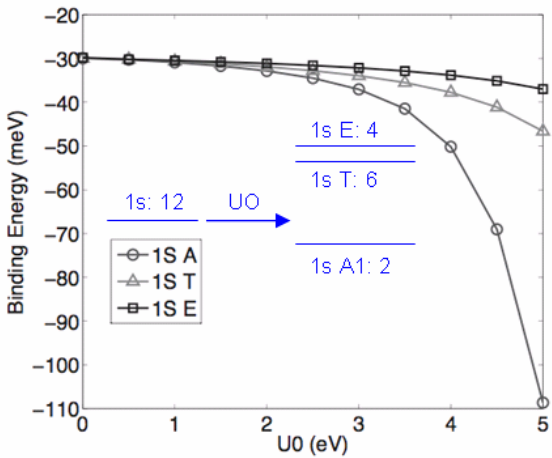

**Figure 31**. Effect of central cell correction $U_0$ on energy splitting. (inset) Group V donor 1s states in Si splitting into 3 components due to valley-orbit interaction.

*Numerical Study of the Valley-Orbit Interaction.* The inset of Figure 31 shows the lowest 6 1s type energy states of a P donor in Si. When valley-orbit coupling is ignored, the six lowest states are degenerate in energy. When Valley-orbit coupling is taken into account, the six fold degenerate states split into a ground state of symmetry A1, a triply degenerate state of symmetry T and a doubly degenerate state of symmetry E. Valley-orbit interaction, which arises due to the deviation of the impurity potential from its bulk-like Coulombic nature, is typically modeled by a correction term for the impurity potential in the vicinity of the donor site. The strength of this core-correcting potential determines the magnitude of the splitting of the six 1s states, and varies from impurity to impurity. Here we consider a core correcting cut-off potential $U_0$ at the donor site, reflecting a global shift of the orbital energies of the impurity. Figure 31 shows how the energy splitting is affected by the strength of $U_0$. For small $U_0$, the six 1s type states are degenerate in energy. As $U_0$ increases in magnitude, we obtain the singlet, triplet and doublet components, as mentioned earlier. Since the triplet (and doublet) states remain degenerate irrespective of $U_0$, we only plot one state of the T (and E) manifold.

This single core-correction term was found to reproduce the donor eigen states within a few meV, and could be adjusted to match the donor ground state binding energy within a μeV. In general however, the tight-binding parameters for Si can only reproduce the full band structure within a limited accuracy. To model high precision donor electronics within a hundredth of a meV, as is needed in many quantum computing applications, additional core-correction terms are required. In semi-empirical Tight-binding, it is only natural to adjust the on-site orbital energies of the P-donor slightly from their Si counterparts to provide this additional correction. In Figure 32, we show variation of the binding energy of the 1s manifold as a function of the on-site orbital energies of the donor site. The four on-site energies considered are Es, Ep, Ed and Es* corresponding to the s, p, d, s* orbitals respectively. The trends in the plots help us establish a recipe for optimizing the core-correction for a donor species to reflect impurity eigen states within the precision of 0.01 meV. For example, if the only donor ground state of A1 symmetry needs to have a higher binding energy, we can adjust either Es or Es*, each of which will push the A1 state deeper in energy without affecting the excited states (Figures 32a and 32d). Figure 32c shows that both the triplet and the doublet state can be adjusted in energy by Ed without affecting the A1 state, while Figure 32b shows that the triplet state alone is affected by Ep. On the other hand, $U_0$ reflects a global shift of all the on-site energies, and can affect all the 1s states, as already shown in Figure 31b. In short there are enough degrees of freedom to empirically adjust the core-correction to obtain very exact eigen values. Once a set of these parameters ($U_0$, Es, Ep, Ed, Es*) is fixed, they can be used for a variety of applications like Stark shift, charge qubits, etc. without any additional modification. To model a generic impurity, it is recommended that $U_0$ be adjusted first so that the ground state binding energy is reproduced accurately. Then one can consider small deviations in a few of these on-site energies to fit the excited states accurately. In most cases, the parameters $U_0$, Ep and Ed can be sufficient for accurate modeling. The plots here were obtained by the tight-binding $sp^3d^5s^*$ model without spin. Clearly, this is an empirical process that does not account fully for the different nature of the impurity atom in a host lattice. Additional mapping which includes the change of the impurity to host coupling matrices could be performed possible based on an input from an *ab initio* method.

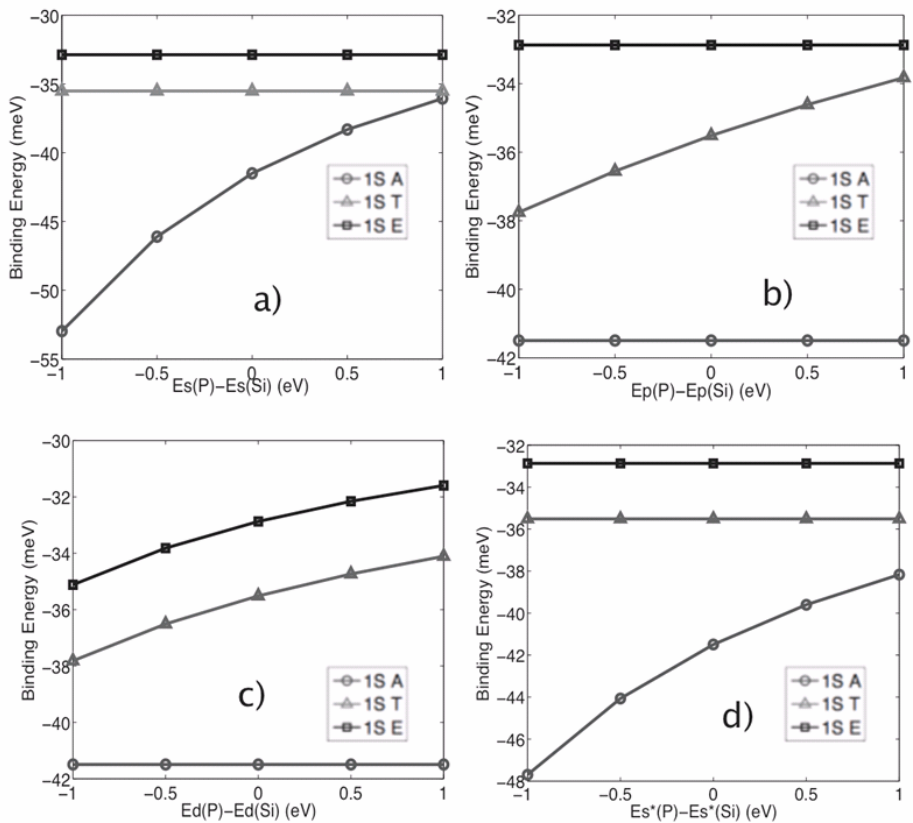

**Figure 32**. Variation of 1s Binding energies with on-site orbital energies. The Triplet (Doublet) states remain degenerate. Hence only 1 triplet (Doublet) is shown.

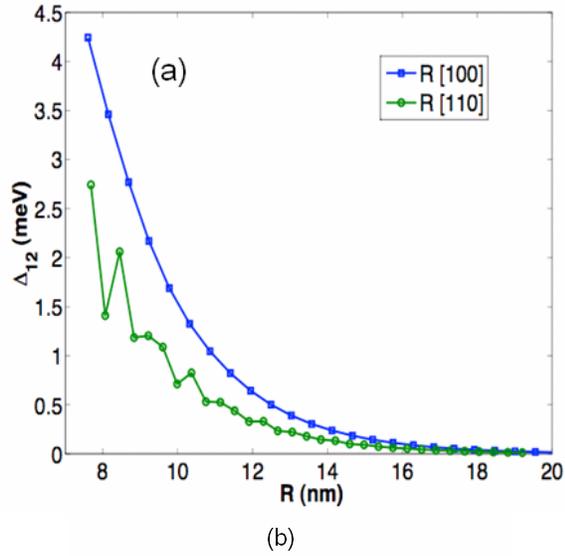

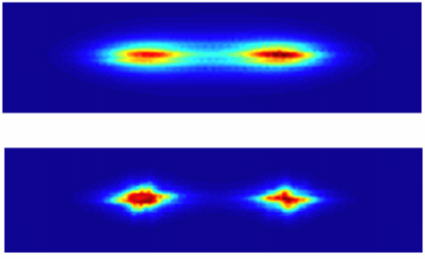

**Figure 33**. (a) Variation of tunnel coupling for a $P_2^+$ system with impurity separation along [100] & [110]. (b) Formation of Molecular states for $P_2^+$.

TABLE V

Comparison of the single donor states relative to the conduction band minima of Si for Lanczos and Block Lanczos algorithms. The Lanczos algorithm fails to capture degenerate states, while Block Lanczos is able to resolve degeneracies at the expense of compute time. The eigenvalues were obtained by the $sp^3d^5s^*$ spin model and shows spin degenerate eigen values as well. The slight deviation of the Eigen values from the experimental values is due to the finite size (i.e. confinement effect) of the simulation domain of 30 nm³.

| EXPERIMENT [5] | LANCZOS | BLOCK LANCZOS (BLOCK SIZE 6) | SYMMETRY |
|---|---|---|---|
| -45.59 | -45.599 | -45.599 | 1s (A1) |
| -45.59 | | -45.599 | 1s (A1) |
| -33.89 | -33.932 | -33.932 | 1s (T) |
| -33.89 | | -33.932 | 1s (T) |
| -33.89 | | -33.930 | 1s (T) |
| -33.89 | | -33.930 | 1s (T) |
| -33.89 | | -33.930 | 1s (T) |
| -33.89 | | -33.930 | 1s (T) |
| -32.58 | -32.67 | -32.670 | 1s (E) |
| -32.58 | | -32.670 | 1s (E) |
| -32.58 | | -32.670 | 1s (E) |
| -32.58 | | -32.670 | 1s (E) |

*Solution Methods—Lanczos & Block Lanczos.* For a realistic simulation involving a few impurities, one needs to consider a lattice size of about 7 million atoms. In atomistic Tight-Binding with a 20 orbital nearest neighbor model, this involves solving a Hamiltonian with 140 million rows and columns. Although this matrix is considerably sparse, solving for interior eigen values occurring near the conduction band poses a difficult problem. Compared to many other algorithms, the parallel Lanczos algorithm for eigen solution has proved very efficient. However, one drawback of the

Lanczos algorithm is its inability to find degenerate and closely clustered eigen values with reliability. A blocked version of Lanczos resolves this problem at the cost of some additional compute time. Since there are many degenerate eigen states present in the unperturbed impurity spectrum, the block Lanczos algorithm was a suitable solution method for the problem outlined here. Table V shows the comparison of eigen states obtained from Lanczos, Block Lanczos, and experimentally established values for single donors in bulk Si. While oscillatory along [110]. This is due to interference between Bloch parts of the impurity wave functions contributed by the Si crystal. These trends are already well established in literature [39] from effective mass theory. The impurity model in TB presented here is able to capture these effects with convenience.

Unlike EMT, the methodology developed here can consider a more extended Bloch structure of the host and incorporate many realistic device effects such as finite device sizes, interfaces under one framework, and

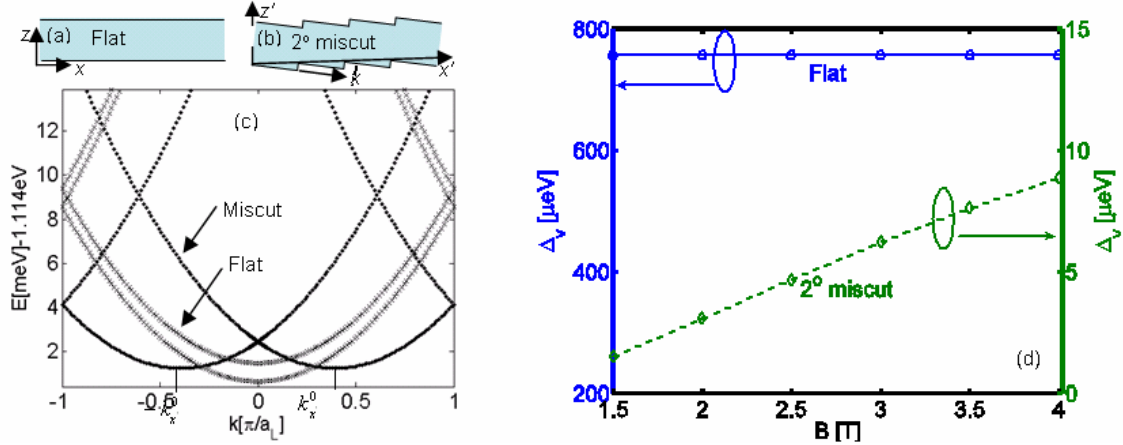

**Figure 34**. (a) Schematic of a Si QW grown on [001] substrate. The crystal symmetry directions are along *x [100]* and *z [001]*. (b) Schematic of a 2º miscut QW unit cell. The unit cell is periodic along *x'* and *y* directions and confined in *z'* direction. Miscut angle is 2º. The step height is *a/4* which corresponds to one atomic layer, where *a* is lattice constant. (c) Band structure of 5.26 nm thick flat/miscut QW along *x/x'* direction. Flat QW shows the presence of two non-degenerate valleys separated by an energy know as VS. Miscut QW shows the presence of two degenerate valleys centered at $\pm k_{x'}^0$. (d) VS in 10nm thick flat (001) and 2º miscut Si QWs. Electric field in z-direction is 9 MV/m.

Lanczos fails to capture the degeneracy of the triplet, doublet and spin states, Block Lanczos resolves all the 12 eigenvalues reliably. The computational system considered here spans a domain of 30.5 nm × 30.5 nm × 30.5 nm and contains about 1.4 million atoms. Closed boundary condition is applied in all three dimensions.

*Typical Application—Donor Based Charge Qubits*. An impurity based charge qubit involves a single electron bound to two ionized P donors in Si. A qubit can be encoded based on the localization of the electron in either of the two impurities [36]. When the Hamiltonian of such a system is solved, a set of bonding and anti-bonding states are obtained from the set of single impurity states. An important parameter in quantum computing applications is the tunnel coupling between the two lowest eigen states. This parameter depends on the separation of the two lowest eigen states of the $P_2^+$ problem, and is sensitive to relative donor placements and gate voltages. Figure 33 shows the tunnel coupling as a function of donor separation along [100] and [110] calculated in tight binding. The tunnel coupling tends to decay, as the impurities are located farther apart. While variation of tunnel coupling is found to be smooth along [100] direction, it is highly is convenient for large scale device simulations. Treatment of such factors enables precise comparison with experimentally measured quantities, as was done in Ref. [82], where the hyperfine stark effect for a P donor was calculated in good agreement with experiment [19], and discrepancies with previous EMT [29] based calculations were resolved. Further work is under way to study CTAP[33] based architectures [37], charge qubits [36][39][52] and investigate donor-interface well hybridization in Si FinFET devices [87][57][20].

### (D) Si on SiGe Quantum Well

Many quantum dot based [30] or impurity based [41] quantum computing architectures are proposed to be fabricated in Si/SiGe heterostures. Since silicon has multiple degenerate values it is critical to engineer these degeneracies out of the system to avoid dephasing of qubits. Miscut substrates (Figure 34b) as opposed to flat substrates (Figure 34a) are often used to ensure uniform growth of Si/SiGe heterostructures. However, a miscut modifies the energy spectrum of a QW. In a flat QW the two degenerate valleys in strained Si split in energy and the valley minima occur at $\pm k_x = 0$. Valley splitting (VS) in a flat QW is a result of interaction among states in bulk z-valleys centered at $k_z = k_0$, where $k_0$ is position of the valley-minimum in strained Si. The energy splitting between these two lowest lying valleys is called as valley-splitting (VS). In quantum computing devices, VS is an important design parameter controlling the electron spin decoherence time [16][17][18]. In a miscut QW lowest lying valleys are degenerate with minima at $\pm k_{x'}^0$ [44]. Thus atomic scale modulation of surface topology leads to very different electronic structures in flat and miscut QWs. As a consequence of this, flat and miscut QWs respond differently to the applied electric and magnetic fields. In the presence of lateral confinement in miscut QW the two degenerate valleys in Figure 34(c) interact and give rise to VS.

Traditional magnetic probe techniques such as Shubnikov de Haas oscillations are used to measure energy spectrum of QWs. Valley and Spin splittings are determined by electron-valley resonance (EVR) [31] and electron-spin resonance (ESR) [26] techniques. In these measurements in plane (lateral) confinement of the Landau-levels is provided by the magnetic field. Figure 34(d) shows the dependence of VS on applied magnetic field in flat and ideal $2^0$ miscut QWs. Ideal miscut QWs refer to the miscut QWs with no step roughness. VS in flat QW is independent of magnetic field because in these QWs VS arises from z-confinement provided by the confining SiGe buffers [44]. In miscut QWs, however, VS is the result of the combined effect of the two confinements, the z-confinement provided by the SiGe buffers and the lateral confinement provided by the applied magnetic field. The two degenerate valleys centered at $\pm k_{x'}^0$ along $x'$ direction in the miscut QWs (Figure 34(c)) interact and split in the presence of magnetic field. At low magnetic fields the dependence of VS in miscut QWs on the applied magnetic field is linear. In calculations of Figure 34(c) and (d) QWs are assumed to be perfect. Disorders such as step roughness and alloy disorder in SiGe buffer which are inherently present is the experiments are completely ignored. As a result calculated VS is nearly an order of magnitude lower than the experimentally measured values (Figure 35(d)).

Miscut substrates undergo reconstruction to reduce the surface free energy which gives rise to the step roughness [105] (Figure 35(b,c)). This type of step roughness disorder is present at the Si/SiGe interface. Another type of disorder in Si/SiGe heterostructures is the random alloy disorder in SiGe buffer. These two disorders are always present in actual QW devices and thus need to be taken into account in VS computations. Schematic of an electronic structure computation domain is shown in Figure 35(a). QWs extend 15 nm along y-direction to take into account the step roughness disorder shown in Figure 35(c). $x'$ confinement due to the magnetic field is incorporated through the Landau gauge $\left(\vec{A} = Bx\hat{y}\right)$. The resulting vector potential $\left(\vec{A}\right)$ is introduced into the tight-binding Hamiltonian trough the gauge invariant Peierl's substitution [10][9][31]. Closed boundary conditions are used in x and z directions while y-direction is assumed to be (quasi-)periodic. The confinement induced by closed boundary conditions in $x'$ direction compete with the magnetic field simulations, the computed VS is higher compared to that of an ideal miscut QW. In these calculations surface roughness model of [6] is used and the uniform biaxial confinement. The lateral extension of the strain and the electronic structure domain is set to 150nm, which is about 7 times larger than the maximum magnetic confinement length in a 2DEG at B = 1.5T ( 21nm). For the magnetic field ranges of 1.5-4T confinement is dominated by the magnetic field and no lateral $x$-confinement effects due to the closed boundary conditions are visible in simulations. Modulation doping in Si/SiGe heterostructures induces built-in electric field. In the simulations performed here constant electric field of 9 MV/m is assumed in the QW growth direction.

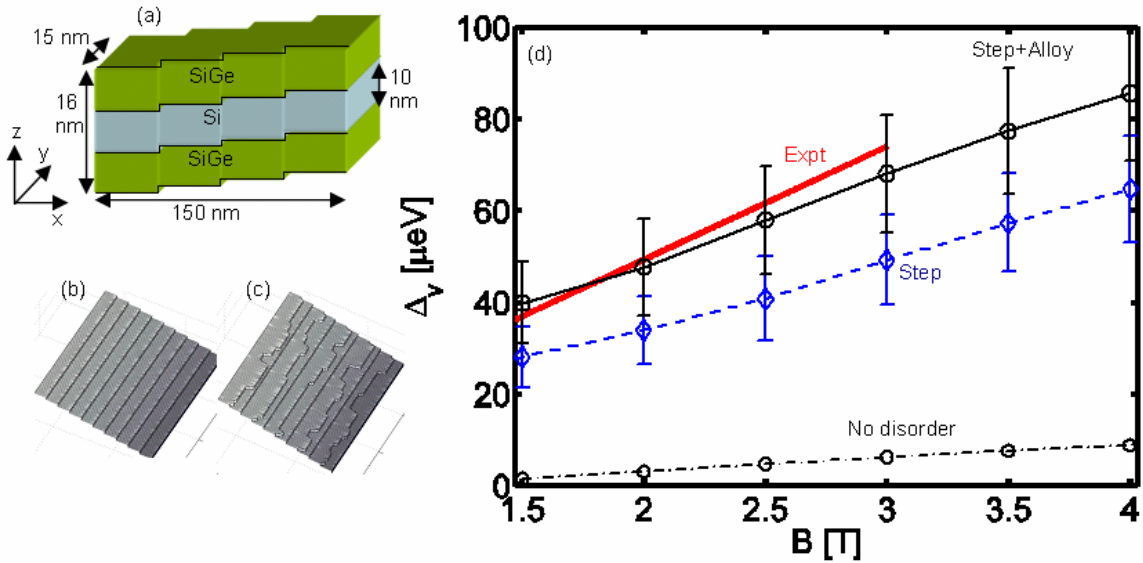

**Figure 35**. (a) Schematics of the simulation domain. (b) Ideal steps on a miscut substrate. (c) Step disorder resulting from the surface reconstruction on the miscut substrate. (d) VS of the first Landau-level in a 10nm thick strained Si QW. The VS labeled as 'No disorder' is shown for comparison and it is same as that of in Fig 1(d). VS increases due to the step-disorder. When alloy-disorder in SiGe buffer is included along with the step disorder the computed VS matches the experimentally measured values. Error bars represent the standard deviation in VS. In the calculations of VS labeled as 'No disorder' and 'Step disorder' uniform biaxial strain of $\varepsilon_\parallel = 0.013$ is assumed.

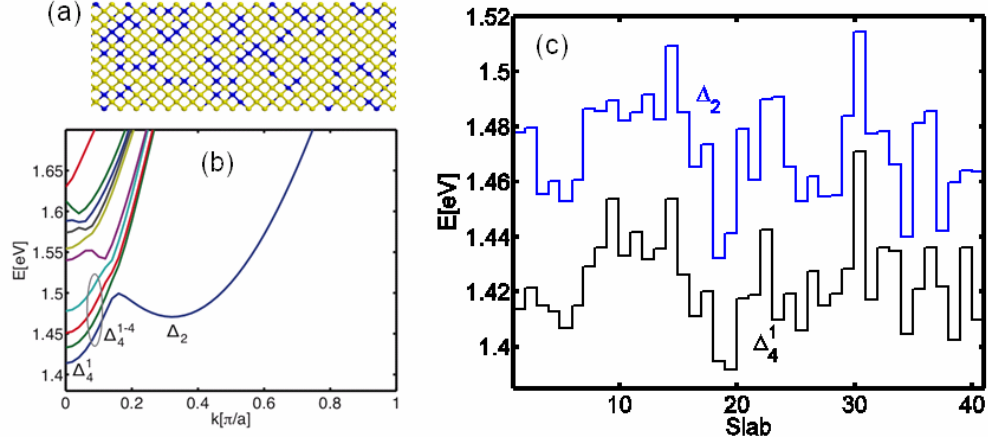

**Figure 36.** (a) Atomistically resolved disorder in the $Si_{0.8}Ge_{0.2}$ 40×6×6 nanowire. (b) Conduction bandstructure of the first slab assuming that the slab is repeated infinitely. $\Delta_4$ valleys are split into four separate bands. $\Delta_2$ valley bands are doubly degenerate. (c) Bandedge minima of lowest energy $\Delta_4$ and $\Delta_2$ valleys along length of the nanowire.

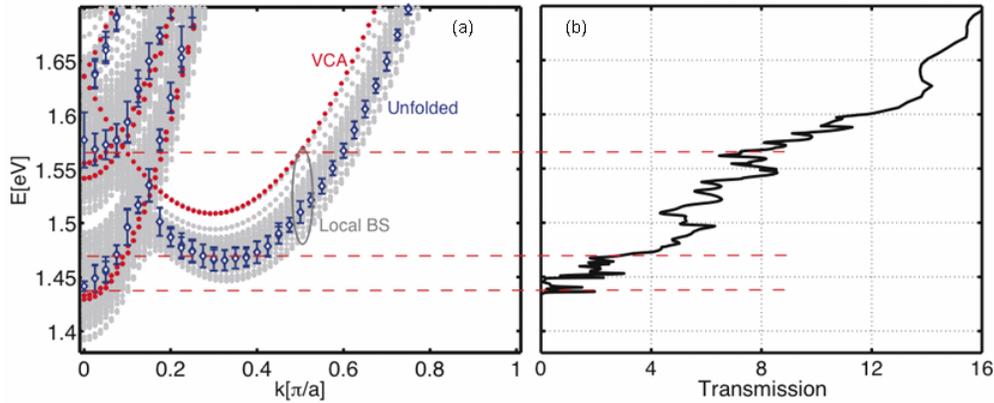

**Figure 37.** (a) Bandstructures of 40×6×6 Si0.8Ge0.2 alloy nanowire in local bandstructure (gray), VCA (red) and zone-unfolding (blue) formulations. (b) Transmission through 40x6x6 wire. Steps in transmission are identified as resulting from new bands appearing in projected bandstructure. Note that atomistic, narrow 1D wires result automatically into 1D localization.

Figure 35(d) shows the computed VS in 2° miscut QWs. VS in ideal miscut QWs is an order of magnitude lower compared to the experimentally measured values. If the step-roughness disorder is included in the simulations, the computed VS is higher compared to that of an ideal miscut QW. In these calculations surface roughness model of [6] is used and the uniform biaxial strain of $\varepsilon_\parallel = 0.013$ which corresponds to $Si_{0.7}Ge_{0.3}$ buffer composition is assumed. This VS, however, is slightly smaller than the experimentally measured VS. This discrepancy can be answered by adding SiGe buffers in the electronic structure simulation domain. 3nm of SiGe buffer is included on top and bottom of the Si QW to take into account the wavefunction penetration into the finite barrier QW buffers. Strain computation domain has the same $x$ and $y$ dimensions as the electronic structure domain. To take into account the long range nature of strain [54] 40nm of SiGe buffer is included on both sides of Si QW. $z$-dimension of the strain domain is 90nm. Valance Force Field (VFF) model of Keating [42] is employed to calculate the relaxed geometries. The VS computed by taking both step and alloy-disorder into account is found to match closely to the experimentally measured values.

The time required to compute the 10 million atom strain calculation on 20 CPUs of an Intel x86-64 dual core linux cluster is about 9 hours. The subsequent 2 million atom electronic structure calculation requires 10 hours.

*(E) SiGe Nanowires*

Semiconductor nanowires are being actively investigated as the potential candidates for the end of the semiconductor technology roadmap devices. They are also attractive for sensing applications due to their high surface-to-volume ration. Several researchers have recently demonstrated the nanowire field-effect transistors (FETs) fabricated from pure elemental or compound semiconductors like Si [24], Ge [33], and GaAs [76] as well as semiconductor alloys like SiGe [45], and their III-V counterparts. For the device design at the nanoscale, it is important to understand and to be able to predict transport properties of nanowires. Atomistic disorder such as alloy disorder, surface roughness and inhomogeneous strain strongly influence the electronic structure and the charge transport in nanoscale devices. To simulate nanodevices tradiational effective mass approaches should be abandoned [98] and more accurate atomistic approaches should be adopted. Here, SiGe alloy nanowires are studied from two different perspectives. First, the electronic structure where bandstructure of a nanowire is obtained by projecting out small cell bands from a supercell eigenspectrum [12][13] and second, the transport where transmission coefficients through the nanowire are calculated using an atomistic wave function (WF) approach [8][64].

SiGe random alloy nanowires have two types of disorders: atom disorder due to random alloying and inhomogeneous strain disorder due to different Si-Si, Ge-Ge, and Si-Ge bond lengths. These disorders break the translational symmetry in semiconductor alloy nanowires. Thus one runs into the problem of choosing a unit cell for the bandstructure calculation. Disorder can be taken into account by simulating larger repeating units (supercells) containing many small cells (Figure 34.). The nanowire bandstructure obtained from the supercell calculation is folded. The one dimensional version of the zone-unfolding method [12][13] is used to project out the approximate eigenspectrum of the nanowire supercell on the small cell Brillouin-Zone. The probability sum rule [12] is used to extract the approximate bandstructure of the alloy nanowire from the projected probabilities. The small cell bandstructure obtained by this method captures the effect of SiGe alloy disorder on the electronic structure.

The nanowire geometry is specified in terms of conventional Zincblende (cubic) unit cells as $n_x \times n_y \times n_z$ where $n_i$ is the number of cubes in direction-$i$. The wire dimensions are 40×6×6 (22.3×3.3×3.3nm) i.e. it is constructed from 40 1×6×6 slabs along [100] crystallographic direction. Figure 36(a) depicts a sliver cut through the center of the SiGe nanowire indicating the atomistically resolved disorder of the wire. Only the central 5nm long portion of this 22nm long wire is shown for visualization purpose. All electronic structure and transport calculations have been done in 20-band $sp^3d^5s^*$ tight-binding model with spin-orbit coupling. The bulk tight-binding and strain Si and Ge parameters are taken from Refs. [15] and [11]. Relaxed wire geometries are calculated from Valance Force Field approach.

The unfolding procedure to compute an approximate bandstructure from the large supercell calculation requires many eigenvectors. In practice these eigenstates are closely spaced in energy and Lanczos algorithm requires about 50000 iterations to resolve 575 states in the energy range of interest. Such calculations require about 5.5 hours on 30 cores of an Intel x86-64 dual core linux cluster machine.

Figure 36(b) shows the conduction bandstructure of the first slab out of 40 slabs along nanowire length. $\Delta_4$ valleys are split into four separate bands while $\Delta_2$ valley bands are doubly degenerate. Local band-edge plots of the lowest $\Delta_4$ and $\Delta_2$ valley minima are shown in Figure 36(c). This so called local bandstructure of each slab is calculated assuming that this slab repeats infinitely along the nanowire. Due to fluctuations in atomic arrangements along the nanowire length one expects to see the different bandstructures for each slab as shown in Figure 37(a). Variations of band-edges along the nanowire length cause reflections which lead to the formation of the localized states and peaks in transmission plots.

The NEMO 3-D team is currently developing with Mathieu Luisier at ETH Zurich a new 3-D quantum transport simulator [64]. Here we show a comparison of a 3-D disordered system transport simulation with a NEMO 3-D electronic structure calculation. The transmission coefficient (Figure 37b) shows the noisy behavior due to random SiGe alloy disorder and inhomogeneous strain disorder in the wire. Steps in the transmission plot can be roughly related to the unfolded bandstructure (Figure 35(a)) from supercell calculations. Four separate $\Delta_4$ valley bands appear as a single band with a finite energy spread in the projected bandstructure. These four bands turn on near 1.44 eV which corresponds to the conduction band transmission turn on. Two $\Delta_2$ valley bands turn on near 1.47 eV which leads to a step in the transmission. 4 more channels due to higher $\Delta_4$ valley sub-bands turn on near 1.57eV. These transmission features can not be related to the conventional virtual crystal approximation (VCA) bandstructure shown in Figure 35(a). Peaks in the transmission plot can be related to the local density of states in the wire [43].

Projected supercell bandstructures and atomistic transport calculations are found to be complimentary and mutually supporting. Both methods provide better instight into the transport through the disordered nanowires.

## VI   SUMMARY AND FUTURE DIRECTIONS

NEMO 3-D is introduced as a versatile, open source *electronic structure* code that can handle device domains relevant for realistic large devices. Realistic devices containing millions of atoms can be computed with reasonably, easily available cluster computers. NEMO 3-D employs a VFF Keating model for strain and the 20-band $sp^3d^5s^*$ empirical tight-binding model for the electronic structure computation. It is released under an open source license and maintained by the NCN, an organization dedicated to develop and deploy advanced nanoelectronic modeling and simulation tools. NEMO 3-D is not limited to research computing alone; A first educational version including visualization capabilities has been released on http://www.nanoHUB.org and has been used by hundreds of users for thousands of simulations. The full version of NEMO3D will soon be available for device engineers, material scientists, educators, and students through the nanoHUB, powered by the NSF Teragrid. Tool documentation, tutorials, and case studies will be posted on nanoHUB as supplemental material. We will generate and deliver tutorials on parallelization and software development through the nanoHUB.

NEMO 3-D demonstrates the capability to model a large variety of relevant, realistically sized nanoelectronic devices. The impact of atomistic strain and piezoelectricity on the electronic structure in dome shaped quantum dots is explored. Under the assumptions of realistic boundary conditions, strain is found to be long-ranged and penetrate around 25 nm into the dot substrate thus stressing the need for using large dimensions of these surrounding layers and at least 3 million atoms in the simulations. The true symmetry of the quantum dots is found to be lower than the geometrical shape symmetry because of the fundamental atomistic nature of the underlying zincblende crystal lattice. Atomistic strain is found to induce further optical polarization anisotropy favoring the [110] direction and pronounced non-degeneracy in the quantum dot excited states, magnitude (few meV) of which depends mainly on the dot size and surrounding material matrix. First order piezoelectric potential, on the other hand, favors the [1$\bar{1}$0] direction, reduces the non-degeneracy in the *P* states and is found to be strong enough to *flip* the optical polarization in certain sized quantum dots [6]. Simulations of QD stacks exemplify the complicated mechanical strain and quantum mechanical interactions on confined electronic states. Molecular states can be observed when the dots are in close proximity. Simulations of SiGe buffered Si QWs indicate the importance of band-to-band interactions that are naturally understood in the NEMO 3-D basis. Valley splitting is computed as a function of magnetic field matching experimental data. Simulations of disordered SiGe alloyed nanowires indicate the critical importance of the treatment of atomistic disorder. Typical approaches of a smoothed out material (VCA) or considerations of bandstructure in just individual slices clearly fail to represent the disordered nanowire physics. A semi-empirical tight binding model for Group V donors in Silicon is presented. The dependence of valley-orbit interaction on on-site cut-off potential and orbital energies is explored. A block based Lanczos algorithm was demonstrated as a robust and reliable method of finding eigenvalues and vectors of the resulting system. The technique outlined here enables high precision modeling of impurity based quantum electronics with relative ease and accuracy.

All these NEMO 3-D calculations underline the importance to represent explicitly the atomistically resolved physical system with a physics based local orbital representation. Such million atom systems result in system sizes of tens of millions and end-to-end 52 million atom simulations representing one *billion* degree of freedom systems were presented. The complexity of the system demands the use of well qualified, tuned, optimized algorithms and modern HPC platforms. Building and maintaining such a code is not a light undertaking and requires a significant group community effort.

Integrated circuit design faces a crisis – the 40 year process of transistor downscaling has led to atomic-scale features, making devices subject to unavoidable manufacturing irregularities at the atomic scale and to heat densities comparable to a nuclear reactor. *A new approach to design that embraces the atomistic, quantum mechanical nature of the constituent materials is necessary to develop more powerful yet energy miserly devices*. We are in the process of developing a general-purpose simulation engine. It will model not only the electronic band structure but also the out-of-equilibrium electron *transport* in realistically extended devices using fully quantum mechanical (QM) models in an atomistic material description containing millions of atoms. The research will enable discovery of new technologies for faster switching, smaller feature size, and reduced heat generation. Using this new approach, designers can directly address questions of quantization and spin, tunneling, phonon interactions, and heat generation. It is widely accepted that the Non-Equilibrium Green Function Formalism (NEGF) QM statistical mechanics theory, in conjunction with an atomistic basis, can answer these questions. It is also widely perceived that the problem is computationally hard to solve. A generalized approach to tri-level parallelism in voltage, energy, and space is highly desired. Another task addresses the bottleneck of calculating open boundary conditions (BCs) for large cross sections for realistically large structures. The BCs can be reused for each voltage point and each charge self-consistent iteration. With a view to achieving these goals, the necessary levels of parallelism to tackle the problem on 200,000+ CPUs have been designed and demonstrated to scale well. Computer scientists and HPC experts embedded in the team will guide the implementation and explore performance, execution reliability, and alternative hardware and algorithms. The new simulation code named OMEN (with non-equilibrium Green function and 3-D atomistic representation) will be an open source project and disseminated through the nanoHUB.

ACKNOWLEDGMENT

The work has been supported by the Indiana 21$^{st}$ Century Fund, Army Research Office, Office of Naval Research, Semiconductor Research Corporation, ARDA, the National Science Foundation. The work described in this publication was carried out in part at the Jet Propulsion Laboratory, California Institute of Technology under a contract with the National Aeronautics and Space Administration. The development of the NEMO 3-D tool involved a large number of individuals at JPL and Purdue, whose work has been cited. Drs. R. Chris Bowen, Fabiano Oyafuso, and Seungwon Lee were key contributors in this large effort at JPL. The authors acknowledge an NSF Teragrid award DMR070032. Access to the Bluegene was made available through the auspices of the Computational Center for Nanotechnology Innovations (CCNI) at Rensselaer Polytechnic Institute. Access to the Oak Ridge National Lab XT3/4 was provided by the National Center for Computational Sciences project. We would also like to thank the Rosen Center for Advanced Computing at Purdue for their support. nanoHUB computational resources were used for part of this work.

BIBLIOGRAPHY

**Primary Literature**

[1] Agnello P D (2002) Process requirements for continued scaling of CMOS—the need and prospects for atomic-level manipulation. IBM J Res & Dev 46:317–338
[2] Ahmed S, Usman M, Heitzinger C, Rahman R, Schliwa A, Klimeck G (2007) Atomistic Simulation of Non-Degeneracy and Optical Polarization Anisotropy in Zincblende Quantum Dots. The 2nd Annual IEEE International Conference on Nano/Micro Engineered and Molecular Systems (IEEE-NEMS), Bangkok, Thailand
[3] Arakawa Y, Sasaki H (1982) Multidimensional quantum well laser and temperature dependence of its threshold current. Appl Phys Lett 40:939
[4] Bae H, Clark S, Haley B, Klimeck G, Korkusinski M, Lee S, Naumov M, Ryu H, Saied F (2007) Electronic structure computations of quantum dots with a billion degrees of freedom. Supercomputing 07, Reno, NV, USA
[5] Bester G, Wu X, Vanderbilt D, Zunger A (2006) Importance of Second-Order Piezoelectric Effects in Zinc-Blende Semiconductors. Phys Rev Lett 96:187602
[6] Bester G, Zunger A (2005) Cylindrically shaped zinc-blende semiconductor quantum dots do not have cylindrical symmetry: Atomistic symmetry, atomic relaxation, and piezoelectric effects. Physical Review B, 71:045318. Also see references therein.
[7] Bowen R, Klimeck G, Roger Lake, Frensley W, Moise T (1997) Quantitative Resonant Tunneling Diode Simulation. J of Appl Phys 81:207
[8] Boykin T, Luisier M, Schenk A, Kharche N, Klimeck G (2007) The electronic structure and transmission characteristics of disordered AlGaAs nanowires. IEEE Trans Nanotechnology 6:43
[9] Boykin T, Vogl P (2001) Dielectric response of molecules in empirical tight-binding theory. Phys Rev B 65:035202
[10] Boykin T, Bowen R, Klimeck G (2001) Electromagnetic coupling and gauge invariance in the empirical tight-binding method. Phys Review B 63:245314
[11] Boykin T, Kharche N, Klimeck G (2007) Brillouin zone unfolding of perfect supercells composed of non-equivalent primitive cells. Phys Rev B 76:035310
[12] Boykin T, Kharche N, Klimeck G, Korkusinski M (2007) Approximate bandstructures of semiconductor alloys from tight-binding supercell calculations. J Phys: Condensed Matter 19:036203
[13] Boykin T, Klimeck G (2005) Practical Application of Zone-Folding Concepts in Tight-Binding. Physical Review B 71:115215
[14] Boykin T, Klimeck G, Bowen R, Oyafuso F (2002) Diagonal parameter shifts due to nearest-neighbor displacements in empirical tight-binding theory. Phys Rev B 66:125207
[15] Boykin T, Klimeck G, Oyafuso F (2004) Valence band effective mass expressions in the sp$^3$d5s$^*$ empirical tight-binding model applied to a new Si and Ge parameterization. Phys Rev B 69:115201
[16] Boykin T, Klimeck G, Eriksson M, Friesen M, Coppersmith S, Allmen P, Oyafuso F, Lee S (2004) Valley splitting in strained Si quantum wells. Applied Physics Letters 84:115
[17] Boykin T, Klimeck G, Eriksson M, Friesen M, Coppersmith S, Allmen P, Oyafuso F, Lee S (2004) Valley splitting in low-density quantum-confined heterostructures studied using tight-binding models. Phys. Rev. B 70:165325
[18] Boykin T, Klimeck G, Allmen P, Lee S, Oyafuso F (2005) Valley-splitting in V-shaped quantum wells. Journal of Applied Physics 97:113702
[19] Bradbury F *et al.* (2006) Stark Tuning of Donor Electron Spins in Silicon. Phys Rev Lett 97:176404
[20] Calder`on, M J, Koiler B, Hu X, Das Sarma S (2006) Quantum Control of Donor Electrons at the Si-SiO2 Interface. Phys Rev Lett 96:096802
[21] Canning A, Wang LW, Williamson A, Zunger A (2000) Parallel Empirical Pseudopotential Electronic Structure Calculations for Million Atom Systems. J of Comp Physics 160:29
[22] Chen P, Piermarocchi C, Sham L (2001) Control of Exciton Dynamics in Nanodots for Quantum Operations. Phys Rev Letters 87:067401

[23] Colinge J P (2004) Multipole-gate SOI MOSFETs. Solid-State Elect 48:897–905
[24] Cui Y, Lauhon L, Gudiksen M, Wang J, Lieber C (2001) Diameter-controlled synthesis of single-crystal silicon Nanowire. Appl Phys Lett 78:2214
[25] Debernardi A *et al.* (2006) Computation of the Stark effect in P impurity states in silicon. Phys Rev B 74:035202
[26] Dobers M, Klitzing K, Schneider J, Weimann G, Ploog K (1998) Electrical Detection of Nuclear Magnetic Resonance in GaAs-Al$_x$Ga$_{1-x}$As Heterostructures. Phys Rev Lett 61:1650
[27] Eriksson M, Friesen M, Coppersmith S, Joynt R, Klein L, Slinker K, Tahan C, Mooney P, Chu J, Koester S (2004) Spin-based quantum dot quantum computing in Silicon. Quantum Information Processing 3:133
[28] Fafard S, Hinzer K, Raymond S, Dion M, Mccaffrey J, Feng Y, Charbonneau S (1996) Red-Emitting Semiconductor Quantum Dot Lasers. Science 22:1350
[29] Friesen M *et al.* (2005) Theory of the Stark Effect for P Donors in Si. Phys Rev Lett 94:186403
[30] Friesen M, Rugheimer P, Savage D, Lagally M, van der Weide D, Joynt R, Eriksson M (2003) Practical design and simulation of silicon-based quantum-dot qubits. Phys Rev B 67:121301
[31] Goswami S, Slinker KA, Friesen M, McGuire LM, Truitt JL, Tahan C, Klein LJ, Chu JO, Mooney PM, van der Weide DW, Joynt R, Coppersmith SN, Eriksson MA (2007) Controllable valley splitting in silicon quantum devices. Nat Phys 3:41
[32] Graf M, Vogl P (1995) Electromagnetic fields and dielectric response in empirical tight-binding theory. Phys Rev B 51:4940
[33] Greentree A, Cole J, Hamilton A, Hollenberg L (2004) Coherent electronic transfer in quantum dot systems using adiabatic passage. Phys. Rev. B 70:235317
[34] Greytak A, Lauhon L, Gudiksen M, Lieber C (2004) Growth and transport properties of complementary germanium Nanowire field-effect transistors. Appl Phys Lett 84:4176
[35] Grundmann M, Stier O, Bimberg D (1995) InAs/GaAs pyramidal quantum dots: Strain distribution, optical phonons, and electronic structure. Phys Rev B 52:11969
[36] Hollenberg L *et al.* (2004) Charge-based quantum computing using single donors in semiconductors. Phys Rev B 69:113301
[37] Hollenberg L *et al.* (2006) Two-dimensional architectures for donor-based quantum computing. Phys Rev B 74:045311
[38] https://www.nanohub.org/simulation_tools/qdot_tool_information
[39] Hu X *et al.* (2005) Charge qubits in semiconductor quantum computer architecture: Tunnel coupling and decoherence. Phys Rev B 71:235332
[40] Jancu J, Scholz R, Beltram F, Bassani F (1998) Empirical spds* tight-binding calculation for cubic semiconductors: General method and material parameters. Phys Rev B 57:6493
[41] Kane B (1998) A Silicon-based Nuclear Spin Quantum Computer. Nature 393:133
[42] Keating P (1966) Effect of Invariance Requirements on the Elastic Strain Energy of Crystals with Application to the Diamond Structure. Phys Rev :145
[43] Kharche N, Luisier M, Boykin T, Klimeck G (2008) Electronic Structure and Transmission Characteristics of SiGe Nanowire. J Comput Electron 7: 350; Klimeck G, Ahmed S, Bae H, Kharche N, Clark S, Haley B, Lee S, Naumov M, Ryu H, Saied F, Prada M, Korkusinski M, Boykin T (2007) Atomistic Simulation of Realistically Sized Nanodevices Using NEMO 3-D: Part I - Models and Benchmarks. IEEE Trans Electron Devices 54:2079; Klimeck G, Ahmed S, Kharche N, Korkusinski M, Usman M, Prada M, Boykin T (2007) Atomistic Simulation of Realistically Sized Nanodevices Using NEMO 3-D: Part II – Applications. IEEE Trans Electron Devices 54:2090
[44] Kharche N, Prada M, Boykin T, and Klimeck G (2007) Valley-splitting in strained Silicon quantum wells modeled with 2 degree miscuts, step disorder, and alloy disorder. *Applied Phys Lett* 90:092109
[45] Kim C, Yang J, Lee H, Jang H, Joa M, Park W, Kim Z, Maeng S (2007) Fabrication of Si$_{1-x}$Ge$_x$ alloy Nanowire field-effect transistors. Appl Phys Lett 91:033104
[46] Klimeck G, Bowen R, Boykin T, Cwik T (2000) sp$^3$s$^*$ Tight-Binding Parameters for Transport Simulations in Compound Semiconductors. Superlattices and Microstructures 27:519-524
[47] Klimeck G, Bowen R, Boykin T, Salazar-Lazaro C, Cwik T, Stoica A (2000) Si tight-binding parameters from genetic algorithm fitting. Superlattices and Microstructures 27:77-88
[48] Klimeck G, Boykin T, Chris R, Lake R, Blanks D, Moise T, Kao Y, Frensley W (1997) Quantitative Simulation of Strained InP-Based Resonant Tunneling Diodes. Proceedings of the 1997 55th IEEE Device Research Conference Digest:92
[49] Klimeck G, Boykin T, Luisier M, Kharche N, Schenk A (2006) A Study of alloyed nanowires from two perspectives: approximate dispersion diagrams and transmission coefficients. proceedings of the 28th International Conference on the Physics of Semiconductors ICPS 2006, Vienna, Austria
[50] Klimeck G, Oyafuso F, Boykin T, Bowen R, Allman P (2002) Development of a Nanoelectronic 3-D (NEMO 3-D) Simulator for Multimillion Atom Simulations and Its Application to Alloyed Quantum Dots. Computer Modeling in Engineering and Science 3:601
[51] Kohn W, Luttinger J (1995) Theory of Donor States in Silicon. Phys Rev 98:915
[52] Koiller B, Hu X, Das Sarma S (2006) Electric-field driven donor-based charge qubits in semiconductors. Phys Rev B 73:045319
[53] Korkusinski M, Klimeck G (2006) Atomistic simulations of long-range strain and spatial asymmetry molecular states of seven quantum dots. Journal of Physics Conference Series 38:75-78
[54] Korkusinski M, Klimeck G, Xu H, Lee S, Goasguen S, Saied F (2005) Atomistic Simulations in Nanostructures Composed of Tens of Millions of Atoms: Importance of long-range Strain Effects in Quantum Dots. Proceedings of 2005 NSTI Conference, Anaheim, CA
[55] Korkusinski M, Saied F, Xu H, Lee S, Sayeed M, Goasguen S, Klimeck G (2005) Large Scale Simulations in Nanostructures with NEMO3-D on Linux Clusters. Linux Cluster Institute Conference, Raleigh, NC
[56] Lanczos C (1950) An Iteration Method for the Solution of the Eigenvalue Problem of Linear Differential and Integral Operators. Journal of Research of the National Bureau of Standards 45
[57] Lansbergen GP, Rahman R, Wellard CJ, Woo I, Caro J, Collaert N, Biesemans S, Klimeck G, Hollenberg LCL, and Rogge S (2008) Gate induced quantum confinement transition of a single dopant atom in a Si FinFET. Nature Physics 4:656
[58] Lazarenkova O, Allmen P, Oyafuso F, Lee S, Klimeck G (2004) Effect of anharmonicity of the strain energy on band offsets in semiconductor nanostructures. Appl Phys Lett 85:4193
[59] Lee S, Kim J, Jönsson L, Wilkins J, Bryant G, Klimeck G (2002) Many-body levels of multiply charged and laser-excited InAs nanocrystals modeled by empirical tight binding. *Phys Rev B* 66:235307

[60] Lee S, Lazarenkova O, Oyafuso F, Allmen P, Klimeck G (2004) Effect of wetting layers on the strain and electronic structure of InAs self-assembled quantum dots. Phys Rev B 70:125307
[61] Lee S, Oyafuso F, Allmen P, Klimeck G (2004) Boundary conditions for the electronic structure of finite-extent, embedded semiconductor nanostructures with empirical tight-binding model. Phys Rev B 69:045316
[62] Liang G, Xiang J, Kharche N, Klimeck G, Lieber C, Lundstrom M (2006) Performance Analysis of a Ge/Si Core/Shell Nanowire Field Effect Transistor. cond-mat 0611226
[63] Loss D, DiVincenzo DP (1998) Quantum computation with quantum dots. Phys Rev A 57:120
[64] Luisier M, Schenk A, Fichtner W, Klimeck G (2006) Atomistic simulation of nanowires in the $sp^3d^5s^*$ tight-binding formalism: From boundary conditions to strain calculations. Phys Rev B 74:205323
[65] Martins A et al. (2004) Electric-field control and adiabatic evolution of shallow donor impurities in silicon. Phys Rev B 69:085320
[66] Maschhoff K, Sorensen D (1996) A portable implementation of ARPACK for distributed memory parallel architectures. Copper Mountain, CO, United States, 9-13 April 1996
[67] Maximov M, Shernyakov Y, Tsatsul'nikov A, Lunev A, Sakharov A, Ustinov V, Egorov A, Zhukov A, Kovsch A, Kop'ev P, Asryan L, Alferov Z, Ledentsov N, Bimberg D, Kosogov A, Werner P (1998) High-power continuous-wave operation of a InGaAs/AlGaAs quantum dot laser. J Appl Phys 83:5561
[68] Michler P, Kiraz A, Becher C, Schoenfeld W, Petroff P, Zhang L, Hu E, Imamoglu A (2000) A Quantum Dot Single-Photon Turnstile Device. Science 290:2282-2285
[69] Moore G (1975) Progress in Digital Integrated Electronics. IEDM Tech Digest pp. 11–13
[70] Moreau E, Robert I, Manin L, Thierry-Mieg V, Gérard J, Abram I (2001) Quantum Cascade of Photons in Semiconductor Quantum Dots. Phys Rev Lett 87:183601
[71] Naumov M, Lee S, Haley B, Bae H, Clark S, Rahman R, Ryu H, Saied F, Klimeck G (2007) Eigenvalue solvers for atomistic simulations of electronic structures with NEMO-3D. 12th International Workshop on Computational Electronics, Amherst, Oct. 7-10, USA.
[72] Oberhuber R, Zandler G, Vogl P (1998) Subband structure and mobility of two–dimensional holes in strained Si/SiGe MOSFET's. Phys Rev B 58:9941–9948
[73] Open Science Grid at http://www.opensciencegrid.org
[74] Oyafuso F, Klimeck G, Allmen P, Boykin T, Bowen R (2003) Strain Effects in large-scale atomistic quantum dot simulations. Phys Stat Sol (b) 239:71
[75] Oyafuso F, Klimeck G, Bowen R, Boykin T, Allmen P (2003) Disorder Induced Broadening in Multimillion Atom Alloyed Quantum Dot Systems. Phys Stat Sol (c) 4:1149
[76] Persson A, Larsson M, Steinström S, Ohlsson B, Samuelson L, Wallenberg L (2004) Solid phase diffusion mechanism for GaAs NW growth. Nat Mater 3:677
[77] Petroff P (2003) Single Quantum Dots: Fundamentals, Applications, and New Concepts. Springer, Berlin
[78] Prada M, Kharche N, Klimeck G (2007) Electronic Structure of Si/InAs Composite Channels. MRS Spring Meeting: April 9 -13, 2007 San Francisco, CA, USA.
[79] Pryor C, Kim J, Wang L, Williamson A, Zunger A (1998) Comparison of two methods for describing the strain profiles in quantum dots. J Apl Phys 83:2548
[80] Qiao W, Mclennan M, Kennell R, Ebert D, Klimeck G (2006) Hub-based Simulation and Graphics Hardware Accelerated Visualization for Nanotechnology Applications. IEEE Transactions on Visualization and Computer Graphics 12:1061–1068
[81] Rahman A, Klimeck G, Lundstrom M (2005) Novel channel materials for ballistic nanoscale MOSFETs bandstructure effects. 2005 IEEE International Electron Devices Meeting, Washington, DC 601-604
[82] Rahman R et al. (2007) High Precision Quantum Control of Single Donor Spins in Silicon. Phys Rev Lett 99:036403
[83] Ramdas A et al. (1981) Spectroscopy of the solid-state analogues of the hydrogen atom: donors and acceptors in semiconductors. Rep Prog Phys 44
[84] Reed M (1993) Quantum Dots. Scientific American 268:118
[85] Reed M, Randall J, Aggarwal R, Matyi R, Moore T, and Wetsel A (1988) Observation of discrete electronic states in a zero-dimensional semiconductor nanostructure. Phys Rev Lett 60:535
[86] Sameh A, Tong Z (2000) The trace minimization method for the symmetric generalized eigenvalue problem. J Comp and Appl Math 123:155-175
[87] Sellier H et al. (2006) Transport Spectroscopy of a Single Dopant in a Gated Silicon Nanowire. Phys Rev Lett 97:206805
[88] Semiconductor Industry Association (2001) International Technology Roadmap for Semiconductors. (http://public.itrs.net/Files/2001ITRS/Home.htm)
[89] Slater J, Koster G (1954) Simplified LCAO Method for the Periodic Potential Problem. Phys Rev 94:1498
[90] Slater J, Koster G (1954) Simplified LCAO Method for the Periodic Potential Problem. Phys Rev 94:1498
[91] Stegner A et al. (2006) Electrical detection of coherent P spin quantum states. Nature Physics 2:835
[92] Stier O, Grundmann M, and Bimberg D (1999) Electronic and optical properties of strained quantum dots modeled by 8-band $k.p$ theory. Phys Rev B, 59:5688
[93] Sze S, May G (2003) Fundamentals of Semiconductor Fabrication. John Wiley and Sons Inc.
[94] TeraGrid at http://www.teragrid.org
[95] Usman M, Ahmed S, Korkusinski M, Heitzinger C, Klimeck G (2006) Strain and electronic structure interactions in realistically scaled quantum dot stacks. proceedings of the 28th International Conference on the Physics of Semiconductors ICPS 2006, Vienna, Austria
[96] Vasileska D, Khan H, Ahmed S (2005) Quantum and Coulomb Effects In Nanodevices. International Journal of Nanoscience 4:305–361
[97] Vrijen R et al. (2000) Electron-spin-resonance transistors for quantum computing in silicon-germanium Heterostructures. Phys Rev A 62:012306.
[98] Wang J, Rahman A, Ghosh A, Klimeck G, Lundstrom M (2005) Performance evaluation of ballistic silicon nanowire transistors with atomic-basis dispersion relations. Applied Physics Letters 86:093113
[99] Wang L, Zunger A (1994) Solving Schrödinger's equation around a desired energy: Application to silicon quantum dots. J of Chem Physics 100:2394
[100] Wellard C, Hollenberg L (2005) Donor electron wave functions for phosphorus in silicon: Beyond effective-mass theory. Phys Rev B 72:085202
[101] Williamson A, Wang L, Zunger A (2000) Theoretical interpretation of the experimental electronic structure of lens-shaped self-assembled InAs/GaAs quantum dots. Phys Rev B 62:12963–12977
[102] Welser J, Hoyt J, Gibbons J (1992) NMOS and PMOS transistors fabricated in strained silicon/relaxed silicon-germanium structures. IEDM Tech Dig 1000–1002
[103] Wong HS (2002) Beyond the conventional transistor. IBM J Res & Dev 46:133–168
[104] http://www.intel.com/cd/software/products/asmo-na/eng/307757.htm
[105] Zandviet H, Elswijk H (1993) Morphology of monatomic step edges on vicinal Si(001). Phys Rev B, 48:14269

[106] Zheng Y, Rivas C, Lake R, Alam K, Boykin T, Klimeck G (2005) Electronic properties of Silicon nanowires. IEEE Tran Elec Dev 52:1097
[107] Zhirnov V V, Cavin III R K, Hutchby J A, Bourianoff G I (2003) Limits to binary logic switch―a Gedanken model. Proc IEEE 91:1934–1939
[108] Zhu W, Han JP, Ma T (2004) Mobility Measurement and Degradation Mechanisms of MOSFETs Made With Ultrathin High-k Dielectrics. IEEE Trans Electron Dev 51:98–105

**Books and Reviews**

[109] Bimberg D, Grundmann M, Ledentsov N (1999) Quantum Dot Heterostructures. Wiley
[110] Datta S (2005) Quantum Transport: Atom to Transistor. Cambridge University Press
[111] Harrison P (2005) Quantum Wells, Wires and Dots: Theoretical and Computational Physics of Semiconductor Nanostructures. Wiley-Interscience